\documentclass[%
 aip,
 jcp,%
 amsmath,amssymb,
reprint,%
]{revtex4-2}

\usepackage{graphicx}% Include figure files
\usepackage{dcolumn}% Align table columns on decimal point
\usepackage{bm}% bold math
\usepackage{hyperref}% add hypertext capabilities
\usepackage[mathlines]{lineno}% Enable numbering of text and display math
\usepackage{xcolor}
\usepackage{csquotes}

\begin{document}

%\preprint{edGER}

\title{Generalized Einstein Relations between Absorption and Emission Spectra in the Electric-Dipole Approximation}%

% Use letters for affiliations, numbers to show equal authorship (if applicable) and to indicate the corresponding author
\author{Jisu Ryu}
\altaffiliation{Present Address: Array Labs Inc., 329 Alma St., Palo Alto, CA 94301, USA}

\author{David M. Jonas}
\email[Correspondence to: ]{david.jonas@colorado.edu}

\affiliation{Department of Chemistry, University of Colorado, Boulder, CO 80309, USA}

\date{\today}% It is always \today, today,
\begin{abstract}
Recently, Ryu et al. showed that two broadened bands connected by a set of four Einstein-coefficient spectra for stimulated and spontaneous single-photon transitions will obey detailed balance at equilibrium if the spectra satisfy generalized Einstein relations. Here, quantum mechanical expressions for Einstein-coefficient spectra are obtained in the electric-dipole approximation using an intramolecular Boltzmann distribution and the quantized field operators in isotropic, dispersive media of Nienhuis and Alkemade. These expressions suggest relationships between Einstein-coefficient spectra and dipole-strength spectra. The electrodynamic relationship between the spectral density for electromagnetic energy and the spectral density for the square of the electric field is developed and used to define dipole-strength spectra in terms of conditional transition probabilities per unit time. These rigorously relate dipole-strength spectra to Einstein-coefficient spectra, thus establishing quantum formulas for dipole-strength spectra and new generalized Einstein relations between dipole-strength spectra. For transitions between two bands, the dipole-strength spectra depend on a single total dipole strength but replace Einstein's degeneracy ratio and transition frequency with a change in standard chemical potential and a single underlying lineshape that is manifested differently in the four spectra. At equilibrium, the relations specify the Stokes' shift between forward and reverse transitions. The relationships between dipole-strength spectra, spontaneous emission spectral densities and stimulated transition cross sections depend on the refractive index, the dielectric constant, and the local field, but not on the derivative of the refractive index. The broadband relationships reduce to known relationships for narrow spectra inside materials and for line spectra in vacuum.
\end{abstract}

\maketitle

\section{INTRODUCTION}
Einstein's quantum-kinetic relations between absorption, stimulated emission, and spontaneous emission for line spectra in vacuum\cite{RN979,RN1457,RN1458,RN2,RN4003} combine finite radiative lifetimes with infinitely narrow line spectra, thus conflicting with the time-energy uncertainty principle. By using a set of four Einstein-coefficient spectra with signed transition frequencies to distinguish absorption from emission, the principle of detailed balance at equilibrium with Planck blackbody radiation leads to thermodynamic-kinetic generalized Einstein relations between broadened spectra that are connected by time-reversal.\cite{RN3964}

The purpose of this paper is to connect Einstein-coefficient spectra with molecular quantum and statistical mechanics in the electric-dipole approximation. The Einstein-coefficient spectra found here depend on conditional probability distributions within the initial band, transition dipole matrix elements between bands, the density of coupled states within the final band, and electromagnetic properties of the molecular environment. The resulting spectra elucidate the role of the refractive index, dielectric constant, and local field in dipole-strength spectra, spontaneous emission spectral densities, transition cross sections, and their relationships.

The generalized Einstein relations between absorption and emission spectra derived from detailed balance for time-reversal invariant systems at thermodynamic equilibrium in ref. \onlinecite{RN3964} are applicable to any process in which single bosonic quasi-particles with energy $h \nu$ are absorbed and emitted. The strongest infrared and optical transitions arise through electric-dipole radiation. Dirac developed an expression for the quantum mechanical conditional transition probability per unit time in order to obtain Einstein's $A$ and $B$ coefficients for electric-dipole transitions under the influence of a quantized radiation field.\cite{RN903,RN3921,RN4110} The derivation of this expression, which Fermi dubbed ``Golden Rule $^{\#}2$'',\cite{RN3979} only becomes valid after a time that allows resolution within the lineshape and only remains valid so long as the initial state is not appreciably depleted.\cite{RN3030,RN287,RN711,RN713} In consequence, its derivation is never valid for a transition from a purely radiatively broadened upper level\cite{RN4109} and the Golden Rule cannot provide a complete derivation of the generalized Einstein relations. Furthermore, since the Golden Rule squares the magnitude of coefficients from first-order perturbation theory to calculate transition probabilities (which are second-order in the perturbation), it imposes an additional ``no-coherence'' restriction on the initial state so that the first-order calculation gives the correct second-order result. Nevertheless, the Golden Rule is often useful beyond its proven validity,\cite{RN3997,RN3905,RN3906} so we use it to  approach the generalized Einstein relations between single-photon absorption and emission in the electric-dipole approximation. The relationship between stimulated emission and spontaneous emission spectra is obtained within the Golden Rule's limits of validity, but the exact relationship between stimulated emission and absorption spectra is only reached through an \textit{ad hoc} hypothesis about broadening by interactions with the thermal reservoir that establish temperature. Assumptions of time-reversal invariance and Planck blackbody radiation allow a detailed-balance derivation without Golden Rule restrictions. Questions about equilibrium with radiation inside dielectric media, the role of the refractive index, dispersion, the electric permittivity, and the local field (in absorption, stimulated emission, and spontaneous emission) are answered using dipole-strength spectra and their relationship to spectral densities for the spontaneous emission rate and the spectra of stimulated transition cross sections.

This paper has four appendices: A relates the positive-frequency spectral density of electromagnetic energy per unit volume in Planck and Einstein's work to the Fourier spectra of the electromagnetic fields; B shows that a pair of linear functionals that obey the generalized Einstein relation can transform a pair of functions that obey the generalized Einstein relation to generate a new pair of functions that obey the generalized Einstein relation; C compares formulas for $B$-coefficient spectra and dipole-strength spectra to literature definitions for the squared magnitude of the transition dipole for line spectra in vacuum; D connects the stimulated transition probabilities from quantum mechanical first-order time-dependent perturbation theory with a classical field to transition probabilities from $B$-coefficient spectra and a spectral density for electromagnetic energy per unit volume.

The formulas in this paper are written in terms of cyclic frequency $\nu$ with dim$[\nu]=cycles/time$ so that the Planck constant has dim$[h]=energy \cdot time/cycles$ and the reduced Planck constant has dim$[\hslash]=energy \cdot time$. Because there are strong arguments that the SI treatment of $cycles$ as dimensionless is flawed,\cite{RN2369,RN2330} balancing factors of $cycles$ are included in formulas. Readers may ignore these factors as the SI does, but keeping track of $cycles$ as a dimension allows conversion from cyclic frequency to angular frequency by replacing [$(cycles),\: \nu,\: h$] with [$2\pi,\: \omega,\: \hslash$] for the equations in this paper. The reverse conversion is not so simple because $2\pi$ and $\hslash$ need not have a frequency origin.\cite{RN3931}  This paper also distinguishes between spectra that can be measured at a single frequency and spectral densities measured by integrating over a range of frequencies. A spectral density is characterized by invariant integrals over corresponding variable ranges and thus transforms under the change of variables theorem as $a^{\nu }(\nu ) = a^{\omega }(\omega = 2 \pi \nu /cycles) (d\omega /d\nu ) = (2 \pi /cycles) a^{\omega }(\omega =2\pi \nu /cycles)$. In contrast, a spectrum is invariant with $b(\nu ) = b(\omega = 2 \pi \nu /cycles)$. Throughout this paper, spectral densities always have a right superscript for the variable(s) to be integrated over, whereas spectra never have right variable superscripts.

\subsection{setting the stage}

We begin by summarizing the approach and results of ref. \onlinecite{RN3964}. Einstein treated the kinetics of single-photon absorption, stimulated emission, and spontaneous emission between infinitely sharp quantum energy levels of a molecule in vacuum.\cite{RN979,RN2} As in ref. \onlinecite{RN3964,RN979}, the word molecule is used here for any finite-sized and bound single-photon absorber; approximate relationships between Einstein-coefficient spectra and dipole-strength spectra to be derived here will further require that the absorber be very small compared to the wavelength of light so only the dipole terms in a multipole expansion contribute appreciably to the Einstein-coefficient spectra and require that the electric-dipole term dominates. The generalized Einstein relations treat the same three types of transitions between energetically broadened bands.\cite{RN3964} If bands overlap in energy, absorption and emission both become possible for each transition direction. Consequently, the generalized Einstein relations replace Einstein's set of three non-negative rate coefficients for an infinitely sharp spectroscopic line with a set of four non-negative Einstein-coefficient spectra for transitions between two bands $R$ and $S$. Energetically broadened bands also require inclusion of intraband transitions with $R = S$, which only have two Einstein-coefficient spectra. For simplicity, the treatment follows Einstein in assuming that the molecule is isotropic or pseudo-isotropic through time-averaging and that the surrounding medium is homogeneous and isotropic. Reference \onlinecite{RN3964} characterized a thermodynamic system by intensive composition, pressure, temperature and chemical potentials related to the Gibbs free energy; in contrast, the treatment here characterizes a quantum system with extensive particle numbers, volume, temperature and chemical potentials related to the Helmholtz free energy. The quantum energy levels in such a canonical ensemble (as defined by Mayer and Goeppert Mayer\cite{RN4006,RN4,RN3991}) are conventionally taken to depend only on the particle numbers and the volume;\cite{RN4006,RN2661,RN2286,RN594} of course, this neglects the system-thermal reservoir interactions that establish temperature and cause fluctuations in the system energy,\cite{RN2661,RN2286,RN594} a delicate point that we will return to after taking the quantum treatment as far as we can.

The conditional transition probabilities per unit time for a molecule in band $S$ to make a single-photon transition to band $R$ are:
\begin{subequations}
\label{eq2}
\begin{flalign}
    \label{eq2b}
    \begin{split}
& { }^{a}\Gamma_{S\to R} (V,T) \equiv \int_{{\kern 1pt}0}^{+\infty } {a_{S\to R}^{\nu } (-\nu ,V,T)d\nu } ,
    \\ & \hspace{17em} \text{ (spontaneous)}
    \end{split} \\
    \label{eq2a}
    \begin{split}
& { }^{b}\Gamma_{S\to R} (u_{+}^{\nu } ;V,T)  \\
& \quad \equiv \int_{{\kern 1pt}0}^{+\infty } 
{[b_{S\to R} (\nu ,V,T)+b_{S\to R} (-\nu ,V,T)]
u_{+}^{\nu } (\nu )d\nu } .
    \\ & \hspace{17em} \text{   (stimulated)}
    \end{split}
\end{flalign}
\end{subequations}
The prior condition for these conditional transition probabilities per unit time is molecular occupation of the initial band $S$. The use of conditional transition probabilities per unit time assumes weak molecule-field coupling.

Equation (\ref{eq2b}) \textit{defines} the Einstein $A$-coefficient spectral density $a_{S \to R}^{\nu}$ for transitions from $S$ to $R$ through spontaneous emission of a photon with cyclic frequency $\nu$ in terms of the conditional transition probability per unit time ${ }^{a}\Gamma_{S\to R}$. Equation (\ref{eq2a}) \textit{defines} the Einstein $B$-coefficient spectrum for stimulated transitions from $S$ to $R$, which may occur through absorption of a photon [$b_{S \to R}(\nu)$ term] or through stimulated emission of a photon [$b_{S \to R}(-\nu)$ term] in terms of the conditional transition probability per unit time ${ }^{b}\Gamma_{S\to R}$ and the positive-frequency spectral density of classical electromagnetic energy per unit volume $u_{+}^{\nu } (\nu )$. This spectral density of radiation is defined so that
\begin{equation}
\label{eq1}
\int_{\nu_{1} }^{\nu_{2} } {u_{+}^{\nu} (\nu )} \;{\kern 1pt}d\nu \equiv U (\nu_{1} \le \left| \nu \right|\le \nu_{2} )
\end{equation}

\noindent is the total classical energy of the radiation per unit volume for cyclic frequencies with absolute value between positive frequencies $\nu_{1} $ and $\nu_{2}$. In a dielectric, the classical electromagnetic energy is the difference between the internal energy with and without the radiation field and thus includes the field-induced kinetic and potential energy of oscillating charge carriers. This classical electromagnetic energy is the sum of photon energies and excludes the divergent quantum zero-point energy of $(1/2)h\nu$ per mode; ``electromagnetic energy'' will always refer to its classical value in this paper.

In a transition from $S$ to $R$, absorption subtracts one photon from a stimulating mode that has one or more photons beforehand; stimulated emission adds one photon to a stimulating mode that has one or more photons beforehand; and spontaneous emission occurs in all directions independently of whether photons are present beforehand. In Einstein-coefficient spectra, a positive \textit{vs.} negative frequency indicates absorption \textit{vs.} emission, which are distinguished by the photon number decreasing \textit{vs.} increasing by one, respectively, and not by the molecular direction of the transition ($S \to R$ \textit{vs.} $R \to S$). The total conditional transition probability per unit time for a single-photon transition from $S$ to $R$ is
\begin{equation}
\label{eq3}
\Gamma_{S\to R} (u_{+}^{\nu } ;V,T)={ }^{b}\Gamma_{S\to R} (u_{+}^{\nu } 
;V,T)+{ }^{a}\Gamma_{S\to R} (V,T).
\end{equation}
The Einstein coefficients are
\begin{subequations}
\label{INTAB}
\begin{flalign}
    \label{INTABa}
    \begin{split}
& A_{S\to R} (V,T)=\int_{0}^{\infty} {a_{S\to R}^{\nu } (-\nu ,V,T)d\nu } ,
    \end{split} \\
    \label{INTABb}
    \begin{split}
& B_{S\to R} (V,T) = \dfrac{1}{cycles} \int_{-\infty}^{+\infty } 
{b_{S\to R} (\nu ,V,T)d\nu }.
    \end{split}
\end{flalign}
\end{subequations}
The same expressions (\ref{eq2} - \ref{INTAB}), with band subscripts interchanged, hold for molecular transitions from $R$ to $S$. These expressions reduce to Einstein's for infinitely narrow lines. The dimensional factor of $cycles$ in Eq. (\ref{INTABb}) keeps the $B$-coefficient independent of the frequency units here (unlike in ref. \onlinecite{RN1457,RN1458}) and gives Einstein's Eq. ($\text{B}'$)\cite{RN979} as ${ }^{b}\Gamma_{s\to r} = (cycles) B_{s\to r} u_{+}^{\nu }(\nu_{sr})$.

The principle of detailed balance between time-reversed processes at equilibrium\cite{RN2662,RN3917,RN3915,RN3918,RN3492} demands that, over any frequency interval, the rate for equilibrium total emission (spontaneous plus stimulated) from $S$ to $R$ must exactly equal the rate for its time-reversed process, which is equilibrium absorption from $R$ to $S$ over the same frequency interval. Let $^{eq}\mathcal{P}_{S} (V,T)$ denote the equilibrium probability for occupation of the initial band $S$. At equilibrium,
\begin{align}
\label{eq4}
\begin{split}
& { }^{eq}\mathcal{P}_{S} [b_{S\to R} (-\nu )u_{\text{BB+}}^{\nu } (\nu )+a_{S\to 
R}^{\nu } (-\nu )]d\nu \\ 
& \quad ={ }^{eq}\mathcal{P}_{R} b_{R\to S} (\nu )u_{\text{BB+}}^{\nu } 
(\nu )d\nu ,
\end{split} 
\end{align}
where $u_{\text{BB+}}^{\nu } (\nu ,V,T)$ is the positive-frequency spectral density of blackbody radiation per unit volume, the frequency interval $d\nu $ can be as small as we like, and the equilibrium composition, volume, and temperature variables have been suppressed. Assuming dilute molecules and that any cavity surrounding the sample is large, the Planck blackbody radiation spectral density may be written in terms of the positive-frequency spectral density of electromagnetic modes per unit volume $G^{\nu }_{+}(\nu ,V,T)$:
\begin{equation}
\label{eq6}
u_{\text{BB+}}^{\nu } (\nu ,V,T)=\frac{h\nu G^{\nu }_{+}(\nu ,V,T)}{\exp (h\nu /k_{\text{B}} T)-1},
\end{equation}
where $h$ is the Planck constant and $k_{\text{B}}$ is the Boltzmann constant.\cite{RN1068} In the limit of large volume, a linear, homogeneous, and isotropic medium that is weakly dispersive and (approximately) transparent has \cite{RN3920,RN3495}
\begin{equation}
\label{eq7}
G^{\nu }_{+}(\nu ,V,T) \approx \dfrac{8\pi \nu^{2}}{(cycles)^3}[n(\nu )]^{2}[\partial (\nu n(\nu ))/ \partial \nu )]\theta (\nu )/c^{3},
\end{equation}
where $c$ is the speed of light in vacuum, the refractive index $n$ depends on frequency, pressure, temperature, composition, \textit{etc.} ($p$, $T$, \textit{etc.} have been suppressed), and $\theta (\nu )$ is a Heaviside unit step function,\cite{RN447} equal to zero for negative frequencies and one for positive frequencies, which restricts $\nu \ge 0$. Solving Eq. (\ref{eq4}) for the equilibrium blackbody radiation and comparing to the Planck distribution in Eq. (\ref{eq6}) immediately suggests the generalized Einstein relations:
\begin{subequations}
\label{eq14}
\begin{equation}
\label{eq14a}
a_{S\to R}^{\nu } (-\nu ,V,T)=h\nu G^{\nu }_{+}(\nu ,V,T)b_{S\to R} (-\nu 
,V,T).
\end{equation}

\begin{flalign}
\begin{split}
\label{eq14b}
& b_{S\to R} (-\nu ,V,T) \\
& \quad \quad = \dfrac{{ }^{eq}\mathcal{P}_{R} (V,T)}{{ }^{eq}\mathcal{P}_{S} 
(V,T)}b_{R\to S} (\nu ,V,T)\exp (-h\nu /k_{\text{B}} T).
\end{split}
\end{flalign}
\end{subequations}
Ref. \onlinecite{RN3964} establishes these relations as the unique solution to Eq. (\ref{eq4}).  The principal additional assumptions are that the transition probability per unit time for emission into a mode of the electromagnetic field that initially contains $l$ photons is proportional to $(l+1)$ and that the energy of a photon is $h\nu$. The relations in Eq. (\ref{eq14}) hold for any molecular statistics; their derivation does not depend on the Boltzmann distribution, quantum mechanics, conservation of energy, or the Bohr frequency condition.

Prior detailed-balance theories of spectra have been restricted with respect to the absorption center frequency $\nu_0$, the linewidth $\Delta \nu$, and the thermal energy $k_{\rm{B}} T$. The molecular results of Einstein for line spectra in vacuum\cite{RN979} require $\Delta \nu \ll \nu_0$ and $\Delta \nu \ll k_{\rm{B}} T/h$. Van Vleck, Weisskopf, and Margenau\cite{RN3500,RN3499} treated transitions in which all photon energies within the lineshape are much less than the thermal energy, ($h(\nu_0 + \Delta \nu /2) \ll k_{\rm{B}} T$). McCumber\cite{RN2267} treated transitions in which the absorption and emission lineshapes are both far above zero frequency ($\Delta \nu [ (1/2) + h \Delta \nu / (k_{\rm{B}} T)] \ll \nu_0$). In contrast, the generalized Einstein relations hold for arbitrary $\nu_0$, $\Delta \nu$, and $k_{\rm{B}} T$.

The generalized Einstein relations require and specify the Stokes' shift between absorption and emission for any homogeneous lineshape with a non-zero linewidth. The well-known relationship\cite{RN2212,RN2415} $(\Delta \nu)^2=(2\lambda)(k_{\rm{B}}T/h)$ between homogeneous Gaussian lineshape variance $(\Delta \nu)^2$, Stokes shift $(2\lambda)$, and thermal energy $k_{\rm{B}}T$ is one case. The generalized Einstein relations further predict that sufficiently broad transitions can have Stokes' shifts so extreme that the normally emissive reverse transition for an absorptive transition can also be absorptive; molecular examples of such extreme Stokes' shifts are known.\cite{RN3964}

Like the Einstein coefficients for line spectra,\cite{RN979,RN3907,RN432} Einstein-coefficient spectra can be used with the Planck blackbody radiation spectral density to show that, at equilibrium, a single molecule in vacuum will have a uniform spatial probability distribution with a Maxwell-Boltzmann velocity probability distribution and that the photons will have a Bose probability distribution. Unlike the demonstrations by Einstein,\cite{RN979} Milne,\cite{RN3907}, and Bothe,\cite{RN432} these results do not rely on any quantum or statistical mechanical assumptions about the molecule.  

With an additional assumption of intramolecular Boltzmann statistics and ideal behavior (ideal gas, ideal mixture,\cite{RN1986} etc.), the band probability ratio in Eq. (\ref{eq14b}) becomes a chemical equilibrium constant $K$, which can be expressed in terms of the change in standard chemical potential, 
\begin{equation}
\label{eqdm}
\dfrac{{ }^{eq}\mathcal{P}_{S} (V,T)}{{ }^{eq}\mathcal{P}_{R} (V,T)}= K(V,T) = \exp 
[-\Delta \mu_{R\to S}^{\circ} (V,T)/k_{\text{B}} T],
\end{equation}
where  
\begin{equation}
\label{eq20}
\Delta \mu_{R\to S}^{\circ} (V,T)=\mu_{S}^{\circ} (V,T)-\mu 
_{R}^{\circ} (V,T)
\end{equation}
is the change in standard chemical potential for the thermal reaction $R\to S$, and $\mu _{S}^{\circ} (V,T)$ is the standard chemical potential\cite{RN2661} for a molecule in band $S$. As usual, the right superscript circle indicates a standard thermodynamic quantity. With Eq. (\ref{eqdm}), the generalized Einstein relation between forward and reverse stimulated spectra in Eq. (\ref{eq14b}) has an ideal system Boltzmann form
\begin{flalign}
\begin{split}
\label{eq21}
& b_{S\to R} (-\nu ,V,T) \\
& =b_{R\to S} (\nu ,V,T)\exp [-(h\nu -\Delta \mu_{R\to S}^{\circ} (V,T))/k_{\text{B}} T].
\end{split}
\end{flalign}
This relationship allows direct spectroscopic determination of thermodynamic standard free energies.\cite{RN2971} For a non-ideal system, the Einstein-coefficient spectra and the equilibrium band populations may depend on system composition variables $x_i$ and the standard chemical potentials must be replaced by composition-dependent formal chemical potentials $\mu _{S}^{\circ'} (V,T, ... x_i, ...)$.\cite{RN3964}

The generalized Einstein relations provide five pairwise relationships between the four Einstein-coefficient spectra for transitions between two bands.  The four spectra are determined by one $B$-coefficient, one change in standard chemical potential, and one underlying lineshape that manifests differently in the four spectra. As an essentially thermodynamic theory, the generalized Einstein relations connect spectra, but do not calculate spectra. This paper obtains results for electric-dipole transitions that connect to quantum and statistical mechanical theories for calculating spectra.

\subsection{approach}

Since the generalized Einstein relations imply a Bose probability distribution for the photons,\cite{RN3964} a self-consistent approach requires a quantized radiation field. Here, Fermi Golden Rule $^{\#}2$ formulas\cite{RN903,RN3645,RN3912,RN3030,RN3979,RN287,RN711,RN713} for Einstein-coefficient spectra are obtained in the electric-dipole approximation using an intramolecular Boltzmann distribution and the quantized field operators in dispersive media.\cite{RN2514,RN4102,RN5} These approximate expressions suggest relationships between Einstein-coefficient spectra and dipole-strength spectra. In Appendix A, the electrodynamic relationship between the spectral density for electromagnetic energy and the spectral density for the square of the electric field is developed. This electrodynamic relationship is used to \textit{define} dipole-strength spectra in terms of conditional transition probabilities per unit time. Comparison to the conditional transition probabilities per unit time that define Einstein-coefficient spectra relates them to dipole-strength spectra in the manner suggested by the Golden Rule formulas and thus derives generalized Einstein relations between dipole-strength spectra from the derivation in ref. \onlinecite{RN3964}.

\section{Golden Rule for spontaneous and stimulated emission}
The first step in deriving the generalized Einstein relations connects the stimulated emission $B$-coefficient spectrum to the spontaneous emission $A$-coefficient spectral density. Since both spectra are rate coefficients, this becomes possible when kinetic rates can be defined. Kinetic rates can only become valid after the density operator within the initial band becomes diagonal in every basis that diagonalizes the thermal equilibrium density operator, so that oscillatory quantum coherence is absent. The conditions required for a master equation are sufficient for valid kinetic rates.\cite{RN713,RN3492} Subsequent relaxation to equilibrium can be understood in terms of populations, which may reach thermal quasi-equilibrium within a band before reaching true thermal equilibrium. In this section, the equations assume only a kinetic rate description, so volume and temperature need not be specified. None of the results in this section depend on the molecular statistics, on molecular statistical mechanics, or on any equilibrium assumption.

From the quantum electrodynamic point of view,\cite{RN713,RN903,RN2315,RN2982} spontaneous and stimulated emission represent two aspects of a single process with a transition probability proportional to  $(l_{\mbox{\footnotesize{\boldmath$\varepsilon$}},\textbf{k}} +1)$ where $l_{\mbox{\footnotesize{\boldmath$\varepsilon$}},\textbf{k}}$ is the number of photons present in the mode before the transition: spontaneous emission arises from the ``1'' and can occur into every electromagnetic mode, regardless of the number of photons present before emission; stimulated emission requires $l_{\mbox{\footnotesize{\boldmath$\varepsilon$}},\textbf{k}} \ge 1$ photons in the electromagnetic mode before emission; both increase the final photon number to $(l_{\mbox{\footnotesize{\boldmath$\varepsilon$}},\textbf{k}} +1)$.  The key difference is that the \textit{total} spontaneous emission rate involves emission into \textit{all} modes (emitting in \textit{all} directions with \textit{all} polarizations), while stimulated emission goes only into  the stimulating mode (emitting in \textit{one} direction with \textit{one} polarization).

Dirac found Einstein's $A$ and $B$ coefficients by a Golden Rule treatment of the electric-dipole interaction as a perturbation of a non-interacting atom $\otimes$ radiation field system.\cite{RN903} Here, the isolated, unperturbed system is (molecule + environment) $\otimes$ radiation field at constant $V$ and the perturbation is again the electric-dipole interaction. The environment is taken large enough so that the molecule's dynamics, in the absence of electric-dipole interaction with the radiation field, can be described by coherent evolution of (molecule + environment) quantum superposition states for as long as desired and their density of states becomes practically continuous for the overall bound system. This large environment delays the inevitable reckoning with energetic fluctuations that are both required by specifying temperature in the canonical ensemble and ultimately necessary to establish a rate --- see section \ref{dv} below. This treatment does not calculate radiative broadening and shifts of the system's energy levels that arise from the electric-dipole interaction.

The single-mode transition probability per unit time is given by an extension of the Golden Rule expression for a transition from a discrete initial state to a final continuum\cite{RN903,RN3645,RN3030} [see ref. \onlinecite{RN711}, Ch. XIII, Eq. (C-31)] to a transition between initial and final continua\cite{RN3912} by prefacing it with averaging over the initial band and photon number, giving
\begin{widetext}
\begin{flalign}
\begin{split}
\label{eq19}
& \Gamma_{S \to R} (\mbox{\boldmath$\varepsilon$},\textbf{k};\mathcal{P}_{{\mbox{\footnotesize \boldmath$\varepsilon$}},{\rm {\bf k}}},P_{S}^{E \eta }) \\ & \quad = \sum\limits_{l} { \mathcal{P}_{\mbox{\footnotesize \boldmath$\varepsilon$},\rm {\bf k}} } 
(l_{{\mbox{\footnotesize \boldmath$\varepsilon$}},{\rm {\bf k}}} ) \left[
\begin{array}{c}
\\[-8pt]
\quad \displaystyle\int_{E' , \eta '} {P_{S}^{E \eta}}(E_{S \eta '}, \eta ') \displaystyle\int_{E, \eta } {\dfrac{1}{{\hslash}^2 }}\left| \left\langle \mbox{\boldmath$\varepsilon$},\textbf{k},l_{\mbox{\boldmath$\varepsilon$},\textbf{k}} \right| \left\langle S,E',\eta ' \right|\hat{{H}}_{int} \left| R,E,\eta \right\rangle \left|  \mbox{\boldmath$\varepsilon$},\textbf{k},(l_{\mbox{\boldmath$\varepsilon$},\textbf{k}}+1) \right\rangle \right|^2 \\[-1pt]
\quad \quad \quad \quad \quad \quad \quad \quad \quad \quad \quad \quad \quad \quad \cdot \rho_{R}^{E \eta }(E_{R \eta}, \eta ) r_{t}(E_{S \eta '} -E_{R \eta } -h\nu_{\textbf{k}} )d\eta dE_{R \eta }d\eta ' dE_{S \eta '} \\[12pt]
+\displaystyle\int_{E' , \eta '} {P_{S}^{E \eta}(E_{S \eta '} ,\eta ')} \displaystyle\int_{E, \eta} {\dfrac{1}{{\hslash}^2 }}\left| \left\langle \mbox{\boldmath$\varepsilon$},\textbf{k},l_{\mbox{\boldmath$\varepsilon$},\textbf{k}} \right| \left\langle S,E',\eta ' \right|\hat{{H}}_{int} \left| R,E,\eta \right\rangle \left|  \mbox{\boldmath$\varepsilon$},\textbf{k},(l_{\mbox{\boldmath$\varepsilon$},\textbf{k}}-1) \right\rangle \right|^2 \\[-1pt]
\quad \quad \quad \quad \quad \quad \quad \quad \quad \quad \quad \quad \quad \quad \cdot \rho_{R}^{E \eta}(E_{R \eta} , \eta ) r_{t}(E_{S \eta '} -E_{R \eta } +h\nu_{\textbf{k}} )d\eta dE_{R \eta }d\eta ' dE_{S \eta '}\\[2pt]
\end{array}
\right] ,
\end{split}
\end{flalign}
\end{widetext}

\noindent where $S$ and $R$ specify the initial and final molecular bands, $\textbf{k}$ specifies the wavevector of the electromagnetic field, $\mbox{\boldmath$\varepsilon$} $ specifies the electric field polarization vector perpendicular to $\textbf{k}$, $\nu_{\textbf{k}}$ is the mode's positive cyclic frequency, and $\mathcal{P}_{{\mbox{\footnotesize \boldmath$\varepsilon$}},{\rm {\bf k}}} (l_{{\mbox{\footnotesize \boldmath$\varepsilon $}},{\rm {\bf k}}}) $
is the probability that mode 
({\mbox{\boldmath$\varepsilon$}},{\rm {\bf k}})
contains 
$l_{\mbox{\footnotesize{\boldmath$\varepsilon$}},\textbf{k}}$ photons before the transition. $P_{S}^{E \eta}$ is the conditional probability density for the initial configuration $\eta '$ with (molecule + environment) energy $E'=E_{S \eta '}$ within band $S$ and $\rho _{R}^{E \eta} $ specifies the density of final (molecule + environment) states with configuration $\eta$ and energy $E=E_{R \eta}$ within band $R$; the right superscripts on the conditional probability density and density of final states indicate they are each spectral densities with respect to $E$ and $\eta$. At constant energy, $\eta '$ and $\eta $ fully specify a complete set of dimensionless (molecule + environment) variables internal to the bands $S$ and $R$, which are averaged over and integrated over, respectively. $\hat{{H}}_{int} $ is the (time-independent) term in the Hamiltonian for the quantized molecule-field interaction; the complete set of variables $\eta$ in the interaction screens out non-interacting final states so that only the density of coupled final states contributes. The initial state energy is $E_{S \eta '} +l_{\mbox{\footnotesize{\boldmath$\varepsilon$}},\textbf{k}} h\nu_{\textbf{k}}$ and the final state energy is $E_{R \eta} +(l_{\mbox{\footnotesize{\boldmath$\varepsilon$}},\textbf{k}} \pm 1) h\nu_{\textbf{k}}$; the final state of the field is fully specified for each emission $(l_{\mbox{\footnotesize{\boldmath$\varepsilon$}},\textbf{k}} + 1)$ or absorption $(l_{\mbox{\footnotesize{\boldmath$\varepsilon$}},\textbf{k}} - 1) $ transition.

The Golden Rule gives
\begin{center}
$r_{t}(E_{S \eta '} - E_{R \eta} \pm h\nu) = t \, \mathrm{sinc}^2[(E_{S \eta '} - E_{R \eta} \pm h\nu)t/2 \hslash]$,
\end{center}
\noindent a symmetric and non-negative function  with $\dim[r] = time$ and an energetic width associated with the time-energy uncertainty principle.\cite{RN3030,RN287,RN711,RN713} $r$ is normalized with respect to integration over cyclic frequency,
\begin{equation}
    \dfrac{1}{cycles} \int_{-\infty}^{\infty} r_{t} (E_{S \eta '} -E_{R \eta } \pm h\nu )d\nu = 1,
\end{equation}
and has only the functional dependencies that are explicitly indicated by subscripts and arguments. In the Golden Rule, the time $t$ cannot be so large that it allows appreciable depletion of the initial state. If that restriction allows times long enough that the rest of the integrand varies much more slowly than $r$, then the subsequent quantum dynamics of a single absorption or emission event exhibits a Golden Rule transition probability that is proportional to time until the initial state becomes appreciably depleted. In the Golden Rule, $r$ is often approximated as proportional to the delta function, $\lim_{t \to \infty} r_t(\Delta E)=2 \pi \hslash \, \delta ^{E} \! (\Delta E)$, for the purpose of simplifying the integral.\cite{RN287,RN711,RN713,RN4110} Since the Golden Rule is not always applicable,\cite{RN4109} we will replace $r_{t}$ in Eq. (\ref{eq19}) by a lineshape $r_{SE'\eta',RE\eta}$ that can depend on the configurations within the initial and final bands. For example, $r_{SE'\eta',RE\eta}$ would be a normalized Lorentzian compatible with the time-energy uncertainty principle in the treatment of longer-time dynamics found in complement $\mathrm{D_{XIII}}$ of ref. \onlinecite{RN711}.

We now substitute the electric-dipole molecule-field interaction into the Golden Rule transition probability per unit time for emission in Eq. (\ref{eq19}). The electric-dipole interaction is 
$\hat{{H}}_{int} 
= -{\rm {\bf \hat{{d}}}}\cdot
{\rm {\bf \hat{{E}}}}_{local} $, where ${\rm {\bf \hat{{d}}}}=\sum {q_{j} {\rm {\bf \hat{{x}}}}_{j} } $ is the vector molecular electric-dipole operator [$q_{j}$ is the charge and ${\rm {\bf \hat{{x}}}}_{j} $ is the vector position operator of electron or nucleus $j$]\cite{RN711} and ${\rm {\bf \hat{{E}}}}_{local} $ is the quantum operator for the local electric field vector that acts on the molecule inside the medium. Dipole interactions depend on the electric field polarization vector $\mbox{\boldmath$\varepsilon$}$ but are not directly dependent on the wavevector $\textbf{k}$, depending on it only indirectly through orthogonality to polarization $\mbox{\boldmath$\varepsilon$}\cdot \textbf{k}=0$ and the frequency $\nu = c |\textbf{k}|/[(2 \pi/cycles) n(\nu)]$.

\begin{widetext}
\begin{flalign}
\begin{split}
\label{LFeq}
& \left| {\left\langle 
{\mbox{\boldmath$\varepsilon$},\textbf{k},l_{\mbox{\footnotesize{\boldmath$\varepsilon$}},\textbf{k}}} \right|\left\langle 
{S,E',\eta '} \right|(-{\rm {\bf \hat{{d}}}}\cdot 
{\rm {\bf \hat{{E}}}}_{local} )\left| {R,E,\eta} \right\rangle \left|
{\mbox{\boldmath$\varepsilon$},\textbf{k},(l_{\mbox{\footnotesize{\boldmath$\varepsilon$}},\textbf{k}} \pm 1)}
\right\rangle } \right|^{2} \\ 
& =\left| {\left\langle {
 \mbox{\boldmath$\varepsilon$},\textbf{k},l_{\mbox{\footnotesize{\boldmath$\varepsilon$}},\textbf{k}}} \right|\left\langle {S,E',\eta '} \right|(-{\rm {\bf \hat{{d}}}}\cdot [\mathbf{f}_{local} 
]\cdot {\rm {\bf \hat{{E}}}})\left| {R,E,\eta} \right\rangle \left| {\mbox{\boldmath$\varepsilon$},\textbf{k},(l_{\mbox{\footnotesize{\boldmath$\varepsilon$}},\textbf{k}} \pm 1)} \right\rangle 
} \right|^{2} \\ 
& \approx \left| {\left\langle 
{S,E',\eta '} \right|({\rm {\bf \hat{{d}}}}'\cdot 
\mbox{\boldmath$\varepsilon$})\left| {R,E,\eta} \right\rangle } \right|^{2} 
f_{local}^{2} \left| {\left\langle {\mbox{\boldmath$\varepsilon$},{\rm {\bf 
k}},l_{\mbox{\footnotesize{\boldmath$\varepsilon$}},\textbf{k}}} \right|\hat{{E}}\left| {\mbox{\boldmath$\varepsilon$},\textbf{k},(l_{\mbox{\footnotesize{\boldmath$\varepsilon$}},\textbf{k}} \pm 1)}\right\rangle } \right|^{2} . 
\end{split}
\end{flalign}
\end{widetext}

In Eq. (\ref{LFeq}), the local electric field ${\rm {\bf \hat{{E}}}}_{local} =[\mathbf{f}_{local} ]\cdot {\rm {\bf \hat{{E}}}}$ at the molecule differs from the macroscopic electric field ${\rm {\bf \hat{{E}}}}$ inside the dielectric. For weak fields, the relationship is a Cartesian tensor transformation ${\rm {\bf \hat{{E}}}}_{local} =[\mathbf{f}_{local} ]\cdot 
{\rm {\bf \hat{{E}}}}$ (second line), which depends on the specific molecular environment. The approximation on the third line assumes the local field can be defined on a length scale that is small compared to the wavelength, yet large compared to molecular dimensions\cite{RN422,RN2902,RN1979,RN4127} so that the local field factor is approximately independent of the molecular band and configuration; this approximate separation leaves specific molecule-environment interactions in the transition dipole.\cite{RN422,RN1979,RN4127} Alternatively, all local field effects can be treated exactly as a transformation of the molecule's electric dipole,\cite{RN3992} ${\rm {\bf \hat{{d}}}}'f_{local} ={\rm {\bf \hat{{d}}}}\cdot [\mathbf{f}_{local} ]$, with
${\rm {\bf \hat{{d}}}}'\cdot {\rm {\bf \hat{{d}}}}'={\rm {\bf \hat{{d}}}}\cdot {\rm {\bf \hat{{d}}}}$ (third line) defining the magnitude of the dimensionless scalar local field factor $f_{local} $. The opposite point of view, in which all local field effects modify the field,\cite{RN4102} is used in gases;\cite{RN344} many studies have examined how this average $f_{local}^{2} $ depends on the solute and on the solvent refractive index when the solute and solvent are in contact.\cite{RN175} In an isotropic and homogeneous medium, the average local electric field is parallel to the macroscopic field (allowing the prime on the dipole to be dropped), and the average $f_{local}^{2} $ is sufficient for linear spectroscopy.  (The local field factor disappears for magnetic-dipole transitions in non-magnetic media.\cite{RN2902}) The transformation to the last line rewrites the macroscopic vector electric field operator as ${\rm {\bf \hat{{E}}}}=\hat{{E}}{\mbox{\boldmath$\varepsilon$}}$, where $\hat{{E}}$ is a scalar operator.

The quantum operator for the macroscopic electric field of traveling waves inside a homogeneous, isotropic, and transparent medium with weak dispersion has been obtained by Nienhuis and Alkemade\cite{RN2514} and by Milonni.\cite{RN4102,RN5} 
\begin{flalign}
\begin{split}
\label{eq22}
& {\rm {\bf \hat{{E}}}}({\rm {\bf x}})=\sum\limits_{
{\mbox{\boldmath$\varepsilon$}},\textbf{k}} {\left[ {\dfrac{n(\nu )}{\epsilon (\nu )[\partial (\nu n(\nu 
))/\partial \nu ]}} \right]_{\nu_{\textbf{k}} }^{1/2} \left({\dfrac{h\nu_{{\rm 
{\bf k}}}}{2V}}\right)}^{1/2}
\\
& \quad \quad \quad \quad \cdot (i\hat{{a}}_{\mbox{\boldmath$\varepsilon$},{\rm 
{\bf k}}} \mbox{\boldmath$\varepsilon$}e^{i\textbf{k}\cdot {\rm {\bf 
x}}}-i\hat{{a}}_{\mbox{\boldmath$\varepsilon$},\textbf{k}}^{\dag } {\mbox{\boldmath$\varepsilon$}}^{\ast }e^{-i\textbf{k}\cdot {\rm {\bf x}}}).
\end{split}
\end{flalign}
\noindent This is Eq. (25) of ref. \onlinecite{RN2514} and Eq. (32) of ref. \onlinecite{RN4102}, rewritten in rationalized Heaviside-Lorentz units (see ref. \onlinecite{RN1580}), where $\epsilon (\nu )$ is the real-valued frequency-dependent relative electric permittivity of the medium, $n(\nu )=[\epsilon (\nu )\mu (\nu )]^{1/2}$ is its real-valued frequency-dependent refractive index, $\mu (\nu )$ is the real-valued frequency-dependent magnetic permeability, $V$ is the volume of the cavity used for electromagnetic field quantization, and $\nu_{\textbf{k}}$ is the cyclic frequency of the field with wavevector \textbf{k}. Traveling waves have been quantized, so frequencies are positive and each component of the wavevector can be either positive or negative.\cite{RN2315}

The photon creation operator has non-zero matrix elements 
\begin{subequations}
\label{eq76}
\begin{equation}
\label{eq76a}
\left\langle {\mbox{\boldmath$\varepsilon$},\textbf{k},(l_{{\mbox{\footnotesize \boldmath$\varepsilon$}},{\rm {\bf k}}}+1)} 
\right|\hat{{a}}_{\mbox{\boldmath$\varepsilon$},\textbf{k}}^{\dag } \left| \mbox{\boldmath$\varepsilon$},\textbf{k},l_{{\mbox{\footnotesize \boldmath$\varepsilon$}},{\rm {\bf k}}}
\right\rangle =\sqrt {l_{{\mbox{\footnotesize \boldmath$\varepsilon$}},{\rm {\bf k}}}+1} , 
\end{equation}
\noindent and the photon annihilation operator has non-zero matrix elements
\begin{equation}
\label{eq76b}
\left\langle {\mbox{\boldmath$\varepsilon$},\textbf{k},(l_{{\mbox{\footnotesize \boldmath$\varepsilon$}},{\rm {\bf k}}}-1)} 
\right|\hat{{a}}_{\mbox{\boldmath$\varepsilon$},\textbf{k}} \left| \mbox{\boldmath$\varepsilon$},\textbf{k},l_{{\mbox{\footnotesize \boldmath$\varepsilon$}},{\rm {\bf k}}}
\right\rangle =\sqrt{l_{{\mbox{\footnotesize \boldmath$\varepsilon$}},{\rm {\bf k}}}} .
\end{equation}
\end{subequations}
\noindent In vacuum, Eq. (\ref{eq22}) reduces to the operator given by Eq. (29) of Appendix 1 in ref. \onlinecite{RN713} and differs from Eq. (13.22) of ref. \onlinecite{RN2315} by a phase convention.

Based on Eq. (\ref{eq22}) and (\ref{eq76}), the matrix elements of the electric field operator, when inserted into the last line of Eq. (\ref{LFeq}), yield a specific factor [in square brackets below] for both absorption and emission.  For emission,
\begin{subequations}
\label{eq23}
\begin{flalign}
\begin{split}
\label{eq23a}
& \left| {\left\langle {\mbox{\boldmath$\varepsilon$},\textbf{k},l_{{\mbox{\footnotesize \boldmath$\varepsilon$}},{\rm {\bf k}}}} 
\right|\hat{{E}}\left| {\mbox{\boldmath$\varepsilon$},\textbf{k},(l_{{\mbox{\footnotesize \boldmath$\varepsilon$}},{\rm {\bf k}}} +1)}\right\rangle } \right|^{2} \\
& \quad \quad \quad \quad =\left[ {\dfrac{n(\nu 
)}{\epsilon (\nu )[\partial(\nu n(\nu ))/\partial\nu ]}} \right]\dfrac{h\nu _{\bf{k}} 
}{V}\dfrac{(l_{{\mbox{\footnotesize \boldmath$\varepsilon$}},{\rm {\bf k}}}+1)}{2}.
\end{split}
\end{flalign}
For absorption,
\begin{flalign}
\begin{split}
\label{eq23b}
& \left| {\left\langle {\mbox{\boldmath$\varepsilon$},\textbf{k},l_{{\mbox{\footnotesize \boldmath$\varepsilon$}},{\rm {\bf k}}}} 
\right|\hat{{E}}\left| {\mbox{\boldmath$\varepsilon$},\textbf{k},(l_{{\mbox{\footnotesize \boldmath$\varepsilon$}},{\rm {\bf k}}} -1)}\right\rangle } \right|^{2} \\
& \quad \quad \quad \quad =\left[ {\dfrac{n(\nu 
)}{\epsilon (\nu )[\partial(\nu n(\nu ))/\partial\nu ]}} \right]\dfrac{h\nu _{\bf{k}} 
}{V}\dfrac{l_{{\mbox{\footnotesize \boldmath$\varepsilon$}},{\rm {\bf k}}}}{2}.
\end{split}
\end{flalign}
\end{subequations}
For a non-dispersive medium, the prefactor in brackets reduces to $[1/\epsilon ]$, yielding the result obtained from Yariv (ref. \onlinecite{RN3778} chapter 5, section 6); the factor in brackets further reduces to one in vacuum (or $[1/\epsilon_{0}]$ in MKS units).

The Golden Rule expression for the proto-$B$-coefficient spectrum $\mathfrak{b}$ for molecular transitions from $S \to R$ that absorb from or emit into a single mode is:
\begin{widetext}
\begin{flalign}
\begin{split}
\label{eqPROTOb}
& \mathfrak{b}_{S \to R}(\mbox{\boldmath$\varepsilon$},\textbf{k},\nu = \pm \nu_{\bf k};P_{S}^{E \eta}) \\
& \quad \quad = \left[ {\dfrac{n(\nu 
)}{\epsilon (\nu )[\partial(\nu n(\nu ))/\partial\nu ]}} \right] \int_{E' ,\eta '} {{P_{S}^{E \eta} (E',\eta ')}}
\int_{E,\eta} {\dfrac{1}{2 {\hslash}^2 }}
\left| {\left\langle 
{S,E',\eta '} \right|({\rm {\bf \hat{{d}}}}'\cdot 
\mbox{\boldmath$\varepsilon$})\left| {R,E,\eta} \right\rangle } \right|^{2} f_{local}^{2} \\
& \quad \quad \quad \quad \quad \quad \quad \quad \quad \quad \quad \quad \quad \: \cdot \rho_{R}^{E \eta}(E , \eta ) r_{SE'\eta',RE\eta}(E_{S \eta '} -E_{R \eta } \pm h\nu_{\textbf{k}} )d\eta dE d\eta ' dE',
\end{split}
\end{flalign}
\end{widetext}
where the + sign applies to absorption of a photon and the $-$ sign applies to emission of a photon. The dispersive dielectric prefactor in brackets and the 2 in the denominator from Eq. (\ref{eq23}) are not part of the electromagnetic energy density (see below) and thus become part of the proto-$B$-coefficient spectrum. The proto-$B$-coefficient spectrum depends on the (possibly non-equilibrium) conditional probability density $P_{S}^{E \eta}$ within the initial band $S$; at thermal equilibrium, $^{eq}P_{S}^{E \eta}$ may be specified by volume and temperature for a canonical ensemble. Inserting the proto-$B$-coefficient and Eq. (\ref{eq23a}) into the first term of Eq. (\ref{eq19}) yields the $S \to R$ total emission probability per unit time into mode $(\mbox{\boldmath$\varepsilon$},\textbf{k})$
\begin{subequations}
\label{eqbetaEMP}
\begin{flalign}
\begin{split}
\label{eqbetaEMPa}
& ^{em}\Gamma_{S \to R} (\mbox{\boldmath$\varepsilon$},\textbf{k};\mathcal{P}_{{\mbox{\footnotesize \boldmath$\varepsilon$}},{\rm {\bf k}}},P_{S}^{E \eta}) \\
& \quad \quad \quad = \mathfrak{b}_{S \to R}(\mbox{\boldmath$\varepsilon$},\textbf{k},\nu = - \nu_{\bf k};P_{S}^{E \eta}) \\
& \quad \quad \quad \quad \cdot \sum\limits_{l} { \mathcal{P}_{{\mbox{\footnotesize \boldmath$\varepsilon$}},{\rm {\bf k}}} }
(l_{{\mbox{\footnotesize \boldmath$\varepsilon$}},{\rm {\bf k}}}) \frac{h\nu_{\bf k}}{V}
(l_{{\mbox{\footnotesize \boldmath$\varepsilon$}},{\rm {\bf k}}}+1).
\end{split}
\end{flalign}
Similarly, Eq. (\ref{eq23b}) yields the absorption probability per unit time from mode $(\mbox{\boldmath$\varepsilon$},\textbf{k})$
\begin{flalign}
\begin{split}
\label{eqbetaEMPb}
& ^{b}\Gamma_{S \to R} (\mbox{\boldmath$\varepsilon$},\textbf{k};\mathcal{P}_{{\mbox{\footnotesize \boldmath$\varepsilon$}},{\rm {\bf k}}},P_{S}^{E \eta}) \\
& \quad \quad \quad = \mathfrak{b}_{S \to R}(\mbox{\boldmath$\varepsilon$},\textbf{k},\nu = + \nu_{\bf k};P_{S}^{E \eta}) \\
& \quad \quad \quad \quad \cdot \sum\limits_{l} { \mathcal{P}_{{\mbox{\footnotesize \boldmath$\varepsilon$}},{\rm {\bf k}}} }
(l_{{\mbox{\footnotesize \boldmath$\varepsilon$}},{\rm {\bf k}}}) \frac{h\nu_{\bf k}}{V}
l_{{\mbox{\footnotesize \boldmath$\varepsilon$}},{\rm {\bf k}}}.
\end{split}
\end{flalign}
\end{subequations}
\noindent The terms proportional to $l_{{\mbox{\footnotesize \boldmath$\varepsilon$}},{\rm {\bf k}}}$ sum to
\begin{equation}
\label{eq9}
\sum\nolimits_{l=0}^\infty  
{ \mathcal{P}_{{\mbox{\footnotesize \boldmath$\varepsilon$}},{\rm {\bf k}}} }(l_{{\mbox{\footnotesize \boldmath$\varepsilon$}},{\rm {\bf k}}}) \dfrac{h\nu_{\bf k}}{V}
l_{{\mbox{\footnotesize \boldmath$\varepsilon$}},{\rm {\bf k}}}
 = u(\mbox{\boldmath$\varepsilon$},{\rm {\bf k}});
\end{equation}

\noindent where $u({\mbox{\boldmath$\varepsilon$}},{\rm {\bf k}})$ is the average classical electromagnetic energy per unit volume from mode $({\mbox{\boldmath$\varepsilon$}},{\rm {\bf k}})$ with positive frequency $\nu_{\bf{k}}$. The terms proportional to 1 can be simplified using the unit sum of photon-number probabilities for each mode,
\begin{equation}
\label{eq9b}
\sum\nolimits_{l=0}^\infty {\mathcal{P}_{{\mbox{\footnotesize \boldmath$\varepsilon$}},{\rm {\bf k}}} 
(l_{{\mbox{\footnotesize \boldmath$\varepsilon$}},{\rm {\bf k}}} )} =1.
\end{equation}
\noindent The resulting conditional probability per unit time for $S \to R$ emission of a photon into a single mode is
\begin{subequations}
\label{eqbetaEMu}
\begin{flalign}
\begin{split}
\label{eqbetaEMua}
& ^{em}\Gamma_{S \to R} (\mbox{\boldmath$\varepsilon$},\textbf{k};u,P_{S}^{E \eta}) \\
& \quad = \mathfrak{b}_{S \to R}(\mbox{\boldmath$\varepsilon$},\textbf{k},\nu = - \nu_{\bf k};P_{S}^{E \eta}) \left[ u({\mbox{\boldmath$\varepsilon$}},{\rm {\bf k}}) + \dfrac{h\nu_{\bf k}}{V} \right],
\end{split}
\end{flalign}
while the corresponding conditional probability per unit time for $S \to R$ absorption of a photon from a single mode is
\begin{flalign}
\begin{split}
\label{eqbetaEMub}
& ^{b}\Gamma_{S \to R} (\mbox{\boldmath$\varepsilon$},\textbf{k};u,P_{S}^{E \eta}) \\
& \quad = \mathfrak{b}_{S \to R}(\mbox{\boldmath$\varepsilon$},\textbf{k},\nu = + \nu_{\bf k};P_{S}^{E \eta}) \left[ u({\mbox{\boldmath$\varepsilon$}},{\rm {\bf k}}) \right].
\end{split}
\end{flalign}
\end{subequations}

In Eq. (\ref{eqbetaEMu}), $u({\mbox{\boldmath$\varepsilon$}},{\rm {\bf k}})$ is the classical electromagnetic energy density per unit volume before the transition from $S \to R$. For a single mode, the part of the conditional emission probability per unit time that is proportional to $u({\mbox{\boldmath$\varepsilon$}},{\rm {\bf k}})$ can be identified with Einstein's stimulated emission and the part proportional to $h\nu_{\bf k}/V$ can be identified with Einstein's spontaneous emission.\cite{RN432} If $u({\mbox{\boldmath$\varepsilon$}},{\rm {\bf k}}) \neq 0$, there is no physically meaningful distinction between stimulated and spontaneous emission into that mode. However, if $u({\mbox{\boldmath$\varepsilon$}},{\rm {\bf k}}) = 0$, one can call emission into that mode spontaneous emission.

In homogeneous and isotropic media, a sum over all modes gives the average positive-frequency spectral density of classical electromagnetic energy per unit volume from the terms proportional to $l_{{\mbox{\footnotesize \boldmath$\varepsilon$}},{\rm {\bf k}}}$,
\begin{equation}
\label{eq11}
\sum\limits_{{\mbox{\footnotesize \boldmath$\varepsilon$}},{\rm {\bf k}}} {u(\mbox{\boldmath$\varepsilon$},{\rm {\bf k}}) } 
=\int {u_{+}^{\nu } (\nu )d\nu } ,
\end{equation}
and the positive-frequency spectral density of electromagnetic modes per unit volume from the terms proportional to 1,
\begin{equation}
\label{eq12}
\sum\limits_{{\mbox{\footnotesize \boldmath$\varepsilon$}},{\rm {\bf k}}} {\frac{h\nu_{{\rm {\bf 
k}}} }{V}=\int {h\nu G^{\nu}_{+}(\nu ,V,T)d\nu } },
\end{equation}
where $u_{+}^{\nu }$ and $G^{\nu}_{+}$ become continuous in the limit of large volume.  The + subscripts indicate that energies or modes per unit volume are obtained by integration over positive-frequency ranges and that these spectral densities are zero for negative frequencies.

For a given frequency, the sum over modes averages over their proto-$B$-coefficient spectra.  For simplicity, we assume the molecule is either isotropic or isotropically oriented on average so that the transition probabilities per unit time are independent of the field polarization and wavevector direction, and 

\begin{equation}
\label{eqDEFb}
b_{S \to R}( \nu;P_{S}^{E \eta}) = {\left\langle 
{\mathfrak{b}_{S \to R}(\mbox{\boldmath$\varepsilon$},\textbf{k},\nu = \pm \nu_{\bf k};P_{S}^{E \eta})}  \right\rangle }_{ \mbox{\footnotesize \boldmath$\varepsilon$},\rm {\bf k} }.
\end{equation}
\noindent Like Eq. (\ref{eqPROTOb}), this relation holds for both absorption $(\nu = +\nu_{\bf k})$ and emission $(\nu = -\nu_{\bf k})$. Summing the single-mode emission probability per unit time in Eq. (\ref{eqbetaEMua}) over all modes and rewriting the sums as frequency integrals using Eqs. (\ref{eq11}) and (\ref{eq12}) gives the conditional probability per unit time for total $S \to R$ emission
\begin{subequations}
\begin{flalign}
\begin{split}
\label{eqbetaEMtot}
& ^{em}\Gamma_{S \to R} (u_{+}^{\nu};P_{S}^{E \eta}) \\
& \quad = \sum\limits_{{\mbox{\footnotesize \boldmath$\varepsilon$}},{\rm {\bf k}}} {^{em}\Gamma_{S \to R} (\mbox{\boldmath$\varepsilon$},\textbf{k};u,P_{S}^{E \eta})} \\
& \quad = \int_{0}^{+\infty}{ b_{S \to R}(-\nu;P_{S}^{E \eta})
[u_{+}^{\nu } (\nu ) + h\nu G^{\nu}_{+}(\nu ,V,T)]d\nu },
\end{split}
\end{flalign}
and the conditional probability per unit time for $S \to R$ absorption
\begin{flalign}
\begin{split}
\label{eqbetaABStot}
& ^{b}\Gamma_{S \to R} (u_{+}^{\nu};P_{S}^{E \eta}) \\
& \quad = \sum\limits_{{\mbox{\footnotesize \boldmath$\varepsilon$}},{\rm {\bf k}}} {^{b}\Gamma_{S \to R} (\mbox{\boldmath$\varepsilon$},\textbf{k};u,P_{S}^{E \eta})} \\
& \quad = \int_{0}^{+\infty}{ b_{S \to R}(\nu;P_{S}^{E \eta})
[u_{+}^{\nu } (\nu )]d\nu }.
\end{split}
\end{flalign}
\end{subequations}

Comparison between the two (negative frequency) emission terms from Eq. (\ref{eq2}) in the total transition probability per unit time given by Eq. (\ref{eq3}) and the two products in Eq.(\ref{eqbetaEMtot}) establishes the generalized Einstein relation between spontaneous and stimulated emission,
\begin{equation}
\label{eqAB}
    a_{S \to R}^{\nu}(-\nu;P_{S}^{E \eta})
 = h\nu G^{\nu}_{+}(\nu ,V,T)b_{S \to R}(-\nu;P_{S}^{E \eta}).
\end{equation}
\noindent The spontaneous emission rate coefficient is given by multiplying the stimulated emission rate coefficient by the number of electromagnetic modes per unit volume and the photon energy (the photon energy appears because $b$ is chosen as the rate coefficient multiplying the electromagnetic energy density rather than the photon density). The physically essential proportionality is the greater number of modes accessible through spontaneous emission. When dispersion is neglected in the density of modes $G^{\nu }_{+}$ from Eq. (\ref{eq7}), Eq. (\ref{eqAB}) reduces to Eq. (7) of ref. \onlinecite{RN3262},  Eq. (3) of ref. \onlinecite{RN379}, and Eq.  (2A) of ref. \onlinecite{RN2971}.

Equation (\ref{eqAB}) holds for any molecular statistics and does not depend on molecular statistical mechanics or any equilibrium assumption. The volume and temperature do not appear as arguments of $a^{\nu}$ and $b$ here because Eq. (\ref{eqAB}) becomes valid as soon as spontaneous and stimulated emission can be described by kinetic rates -- the distribution $P_{S}^{E \eta}$ in the initial band need not be at equilibrium. Reviewing the above derivation of Eq. (\ref{eqAB}), \textit{none} of the molecular details that went into the stimulated emission Einstein $B$-coefficient spectrum mattered because they went into the spontaneous emission Einstein $A$-coefficient spectral density in the same way. Provided that the dynamics within bands $R$ and $S$ give a sufficiently broad lineshape for the Golden Rule to apply, the form of the function $r$ in Eqs. (\ref{eq19}) and (\ref{eqPROTOb}) is irrelevant to derivation of the generalized Einstein relation between spontaneous and stimulated emission and Eqs. (\ref{eqAB}) and (\ref{eq14a}) can be derived from the Golden Rule within its limits of validity.

Equation (\ref{eqAB}) also holds at thermal quasi-equilibrium with volume $V$ and temperature $T$ as additional arguments for both $a$ and $b$. Thermal quasi-equilibrium requires that the conditional probability density for configurations $\eta '$ within the initial band $S$ take on its equilibrium distribution $P_{S}^{E \eta} = {^{eq}P}_{S}^{E \eta}(V,T)$, but allows a non-equilibrium prior probability $\mathcal{P}_S \neq {^{eq}\mathcal{P}_S (V,T)}$ and an arbitrary distribution within the final band $R$. Thermal quasi-equilibrium is the condition for the existence of quasi-Fermi levels in solid state physics.\cite{RN1795} After thermal quasi-equilibrium is established, the non-equilibrium emission lineshape becomes independent of the excitation wavelength. Equation (\ref{eq14a}) is a special case of Eq. (\ref{eqAB}) under the condition of thermal quasi-equilibrium within the initial band $S$.

\section{\label{secFR} Forward and Reverse Transitions}
\subsection{equilibrium and the Golden Rule approach}
We now use an assumption of thermal quasi-equilibrium within the initial band and a Boltzmann distribution  to connect the stimulated forward and reverse Einstein $B$-coefficient spectra in Eq. (\ref{eq14b}). Since we are treating a single molecule, there is no need to assume Boltzmann statistics and the results of this section apply to any molecular statistics. At thermodynamic equilibrium, the standard chemical potential for band $S$ is given by the usual expression
\begin{equation}
\label{eqSCP}
\mu_{S}^{\circ} (V,T)=-k_{\text{B}} T \ln[Z_S(V,T)],
\end{equation}
where the single-molecule partition function for band $S$ is
\begin{flalign}
\begin{split}
\label{eqzS}
& Z_S(V,T) \\
& \quad =\displaystyle\int_{E', \eta '}
 \rho_{S}^{E \eta}(E_{S \eta '} , \eta ')
 \exp(-E_{S \eta '} / k_{\text{B}} T )
 d \eta' dE_{S \eta '} \; .
\end{split}
\end{flalign}
This integral of the (molecule + environment) density of states within band $S$ is Boltzmann weighted by (molecule + environment) energies $E_{S \eta '}$; the energies for all bands must be referenced to a common zero. The statistical mechanical equilibrium conditional probability density in the initial band is given by
\begin{equation}
\label{eqP}
^{eq}P_{S}^{E \eta}(E_{S \eta '} , \eta ') = \dfrac { \rho_{S}^{E \eta}(E_{S \eta '} , \eta ') \exp[-E_{S \eta '} / k_{\text{B}} T ] } {\exp[ - \mu ^\circ _S (V,T) / k_{\text{B}} T ]},
\end{equation}
which also holds at quasi-equilibrium. Substituting these equilibrium conditional probability densities into the proto-$B$-coefficients given by Eq. (\ref{eqPROTOb}) and eliminating the standard chemical potentials with Eqs. (\ref{eqdm}) and (\ref{eq20}) gives the ratio
\begin{widetext}
\begin{flalign}
\begin{split}
\label{eqPROTObRATIO}
& \dfrac{^{eq}\mathcal{P}_R \mathfrak{b}_{R \to S}(\mbox{\boldmath$\varepsilon$},\textbf{k},\nu = + \nu_{\bf k}; {^{eq}P_{R}^{E \eta})}}{^{eq}\mathcal{P}_S \mathfrak{b}_{S \to R}(\mbox{\boldmath$\varepsilon$},\textbf{k},\nu = - \nu_{\bf k}; {^{eq}P_{S}^{E \eta})}} \\
& \quad = \dfrac
{\left[
\begin{array}{lc}
\displaystyle\int_{E, \eta } \displaystyle\int_{E', \eta '} & {\rho_{R}^{E \eta}(E_{R \eta } , \eta ) \exp(-E_{R \eta} / k_{\text{B}} T )}
\left| {\left\langle 
{R,E,\eta} \right|({\rm {\bf \hat{{d}}}}'\cdot 
\mbox{\boldmath$\varepsilon$})\left| {S,E',\eta '} \right\rangle } \right|^{2} \rho_{S}^{E \eta}(E_{S \eta '} , \eta ') \\
& \cdot r_{RE\eta,SE'\eta'}(E_{R \eta} -E_{S \eta '} + h\nu_{\textbf{k}} , T)d\eta ' dE_{S \eta '}d \eta dE_{R \eta } 
\end{array}
\right] }
{\left[
\begin{array}{lc}
\displaystyle\int_{E' ,\eta '} \displaystyle\int_{E, \eta} & {\rho_{S}^{E \eta}(E_{S \eta '} , \eta ') \exp(-E_{S \eta '} / k_{\text{B}} T )} \left| {\left\langle {S,E',\eta '} \right|({\rm {\bf \hat{{d}}}}'\cdot \mbox{\boldmath$\varepsilon$})\left| {R,E,\eta} \right\rangle } \right|^{2} \rho_{R}^{E \eta}(E_{R \eta} , \eta ) \\
& \cdot r_{SE'\eta',RE\eta}(E_{S \eta '} -E_{R \eta } - h\nu_{\textbf{k}} ,T)d \eta dE_{R \eta } d \eta ' dE_{S \eta '} 
\end{array}
\right] } .
\end{split}
\end{flalign}
\end{widetext}

\noindent Invoking the Hermitian property of the quantum matrix elements, the only sources of asymmetry between numerator and denominator on the right-hand side of Eq. (\ref{eqPROTObRATIO}) are the Boltzmann factors and the functions $r_{RE\eta,SE'\eta'}$ and $r_{SE'\eta',RE\eta}$, in which both the indices and the argument signs are reversed. Based on this observation, the forward and reverse proto-$B$-coefficient spectra would satisfy the generalized Einstein relation in Eq. (\ref{eq14b}) if
\begin{flalign}
\begin{split}
\label{rGER}
& {r_{SE'\eta',RE\eta}(E_{S \eta '} -E_{R \eta } - h\nu_{\textbf{k}} ,T)} \\
& \quad \quad \mathop =\limits^{?} {r_{RE\eta,SE'\eta'}(E_{R \eta} -E_{S \eta '} + h\nu_{\textbf{k}} ,T)} \\ 
& \quad \quad \quad \cdot \exp[(E_{S \eta '} -E_{R \eta } - h\nu_{\textbf{k}})/k_{\text{B}} T]
\end{split}
\end{flalign}
were to hold. Especially as an exact equality between a normalized pair of non-negative functions $r$, the ``question-marked'' Eq. (\ref{rGER}) is an \textit{ad hoc} hypothesis proposed in order to satisfy the generalized Einstein relation. The temperature was added as an argument of $r$ in Eq. (\ref{eqPROTObRATIO}) because Eq. (\ref{rGER}) requires that one or both of the $r$ depend on temperature for any non-zero width. Equality (\ref{rGER}) contains only system energies; if one were to approximate the Golden Rule function $r_t$ as an energy-conserving delta function for the total (system + thermal reservoir) energies (as in $\S$3.3 of ref. \onlinecite{RN4117}), the exponential factor in Eq. (\ref{rGER}) could be understood as quantifying the relative availability of thermal reservoir energy for supplying or removing the difference between system energies in the forward \textit{vs.} reverse transitions. Equality (\ref{rGER}) would make the integrands in the numerator and denominator on the right-hand side of Eq. (\ref{eqPROTObRATIO}) identical except for a factor of $\exp(-h \nu_{\textbf{k}}/k_{\text{B}} T)$, which could be pulled outside the integral in the denominator to show that the ratio obeys the generalized Einstein relation:
\begin{flalign}
\begin{split}
\label{eqPROTObGER}
& {^{eq}\mathcal{P}_S \mathfrak{b}_{S \to R}(\mbox{\boldmath$\varepsilon$},\textbf{k},\nu = - \nu_{\bf k}; {^{eq}P_{S}^{E \eta})}} \\
& \quad \mathop =\limits^{?}  {^{eq}\mathcal{P}_R \mathfrak{b}_{R \to S}(\mbox{\boldmath$\varepsilon$},\textbf{k},\nu = + \nu_{\bf k}; {^{eq}P_{R}^{E \eta})}}
\exp[-h\nu_{\bf k} / k_{\text{B}}T].
\end{split}
\end{flalign}
The ``question-marked'' Eq. (\ref{eqPROTObGER}) depends upon Eq. (\ref{rGER}). Equality (\ref{rGER}) is sufficient to obtain the single-mode generalized Einstein relation of Eq. (\ref{eqPROTObGER}) and would thus give the less-detailed generalized Einstein relations of Eq. (\ref{eq14}) in homogeneous, isotropic, and time-reversal invariant systems, but it has not been established as necessary. The two single mode processes with rates connected by Eq. (\ref{eqPROTObGER}) are not time-reversed with respect to each other, but might be called ``conjugate processes'' because of the complex-conjugate quantum matrix elements in their proto-$B$-coefficients. 

To clarify the ratio of proto-$B$-coefficient spectra and the role of Eq. (\ref{rGER}), we note that the delta function approximation\cite{RN4110} to the Golden Rule
\begin{flalign}
\begin{split}
\label{rdelta}
& r_{SE'\eta',RE\eta}(E_{S \eta '} -E_{R \eta } - h\nu_{\textbf{k}} ) \\
& \quad \quad = r_{RE\eta,SE'\eta'}(E_{R \eta } -E_{S \eta '} + h\nu_{\textbf{k}} ) \\
& \quad \quad \approx 2 \pi \hslash \,\delta^{E} \! (E_{S \eta '} -E_{R \eta } - h\nu_{\textbf{k}} ),
\end{split}
\end{flalign}
is the narrowest member of the class of paired linear functionals $r$ obeying Eq. (\ref{rGER}) discussed in Appendix B. Equality (\ref{rGER}) does not hold for the non-zero width function $r_t$ in the Golden Rule, nor does it hold for any even symmetric function, such as a Lorentzian. Even symmetry functions of $(E_{S \eta '} -E_{R \eta } - h\nu_{\textbf{k}})$ with a non-zero width do not satisfy detailed balance with Planck blackbody radiation in Eq. (\ref{eq4}). 

\subsection{isotropic averaging}
We now relate the isotropic Einstein-coefficient spectra to isotropic dipole-strength spectra so that generalized Einstein relations can be formulated for dipole-strength spectra. Substituting Eq. (\ref{eqPROTOb}) into Eq. (\ref{eqDEFb}), we obtain the isotropic Einstein $B$-coefficient spectrum
\begin{widetext}
\begin{flalign}
\begin{split}
\label{eq25}
& b_{S\to R} (\nu = \pm \nu_{\textbf{k}}, V, T) \\
& =\left[ {\dfrac{n(\nu )}{\epsilon (\nu )[\partial(\nu n(\nu ))/\partial\nu ]}} \right] \dfrac{1}{2\hslash^{2}}
\biggl \langle { f_{local}^{2}
\displaystyle{\int}_{E',\eta'} {^{eq}P_{S}^{E \eta} (E',\eta ')} \displaystyle{\int}_{E,\eta} 
{\left| {\left\langle {S,E',\eta '}\right|({\rm {\bf \hat{d}} '}\cdot \mbox{\boldmath$\varepsilon$})\left| {R,E,\eta}\right\rangle } 
\right|^{2} } } \\
& \quad \quad \quad \quad \quad \quad \quad \quad \quad \quad \quad \quad \quad \quad { \cdot r_{SE'\eta',RE\eta}(E_{S \eta '} -E_{R \eta } \pm h\nu_{\textbf{k}} ,T) \rho_{R}^{E \eta} (E, \eta ) d\eta dE d\eta' dE'} \biggr \rangle 
_{\mbox{\boldmath$\varepsilon$},\bf{k}}
\end{split}
\end{flalign}
for an electric-dipole transition. The angular brackets enclose all factors that may depend on the specific molecule.

The averages over $\mbox{\boldmath$\varepsilon$}$ and $\bf{k}$ can be brought inside both integrals for integrating and averaging over intramolecular configuration variables. For dipole radiation, an average over wave-vector directions is not needed and an average over any three perpendicular laboratory frame polarization vectors $\mbox{\boldmath$\varepsilon$}$ is sufficient to generate the isotropic light average;
\begin{equation}
\label{eq26}
\begin{array}{l}
 \left\langle {\left| {\left\langle {S, E',\eta '} 
\right|({\rm {\bf \hat{{d}}'}}\cdot \mbox{\boldmath$\varepsilon$})\left| {R, E, \eta}
\right\rangle } \right|^{2} } \right\rangle_{\mbox{\boldmath$\varepsilon$}}
=\dfrac{1}{3} \cdot \sum\limits_{\mbox{\boldmath$\varepsilon$}=X,Y,Z} {\left| 
{\left\langle {S, E', \eta '} \right|({\rm {\bf 
\hat{{d}}'}}\cdot \mbox{\boldmath$\varepsilon$})\left| {R, E, \eta} \right\rangle } 
\right|^{2} } 
\equiv \dfrac{1}{3}\left| {{\rm {\bf d}'}_{S E'\eta ', R E \eta } } 
\right|^{2} \\ 
\end{array}
\end{equation}
this yields the factor of (1/3) and sum after the equal sign. The equivalence on the far right defines an abbreviation for the sum. Inserting this abbreviation into Eq. (\ref{eq25}) yields the Golden Rule expression for the dipole-strength spectrum of the transition $S \to R$ as 
\begin{flalign}
\begin{split}
\label{eq27}
& d_{S\to R}^{2} (\nu, V, T) \\
& \quad = \int_{E' , \eta'} 
{ {^{eq}P_{S}^{E \eta} (E',\eta ')} \int_{E,\eta} {\left| {{\rm {\bf d}'}_{S E'\eta ', R E \eta } } \right|^{2}
r_{SE'\eta',RE\eta}(E_{S \eta '} -E_{R \eta } + h\nu , T)
\rho_{R}^{E \eta } (E, \eta ) d\eta dE d\eta' dE } }.
\end{split}
\end{flalign}
\end{widetext}
Since $\eta $ and $\eta '$ are dimensionless variables and $\dim [r] = time$, the dipole-strength spectrum has $\dim [d_{S\to R}^{2} (\nu )]=(charge\cdot length)^{2}\cdot time$ and is independent of the frequency units. Equation (\ref{eq27}) is the key quantum formula for the dipole- strength spectrum, which is related to literature definitions of the squared magnitude of the transition dipole between levels in Appendix C.

The Golden Rule expressions for $b_{S \to R}$ and $d_{S \to R}^2$ obey 
\begin{equation}
\label{eq28}
\begin{array}{l}
 b_{S\to R} (\nu )=f_{local}^{2}(\nu)\left[ {\dfrac{ n(\nu )}{\epsilon (\nu )[\partial(\nu n(\nu ))/\partial \nu ]}} \right]\dfrac{1}{6\hslash^{2}}d_{S\to R}^{2} (\nu ). \\ 
 \end{array}
\end{equation}
Neglecting the local field factor, this reduces to the medium dependence for line spectra in ref. \onlinecite{RN2514}. Alternatively, specializing to a Lorentz-Lorenz local field and non-magnetic media reduces it to the refractive-index dependence for line spectra in ref. \onlinecite{RN4102}. Specializing even further to non-dispersive media reduces this to the refractive index dependence for line spectra in ref. \onlinecite{RN3993}.

\section{dipole-strength spectra}
Rewriting the approximate Golden Rule expression for the conditional transition probability per unit time for a molecule at spatial position $\rm {\bf x}$ in band $S$ to make a stimulated single-photon, electric-dipole transition to band $R$ in Eq. (\ref{eq2a}) by substituting for $b$ using Eq. (\ref{eq28}) and for $u_{+}^{\nu}$ using Eq. (\ref{sdu2sdE}) gives
\begin{flalign}
\begin{split}
\label{eq2d}
& { }^{b}\Gamma_{S\to R} (({{\rm {\bf E} \cdot {\bf E}}})^{\nu}_{+} ;V,T)  \\
& \equiv \dfrac{1}{6 \hslash^2}
\int_{0}^{+\infty }
f_{local}^{2}(\nu) 
[d_{S\to R}^{2} (\nu ,V,T)
  + d_{S\to R}^{2} (-\nu ,V,T)]\\
& \quad \quad \quad \quad \quad \quad \quad \cdot ({{\rm {\bf E} \cdot {\bf E}}})^{\nu}_{+} ( {\rm {\bf x}}, \nu) d\nu , \\
\end{split}
\end{flalign}
We use Eq. (\ref{eq2d}) to \textit{define} the dipole-strength spectrum in the same way as Eq. (\ref{eq2a}) defines the $B$-coefficient spectrum. The dipole-strength spectra defined by Eq. (\ref{eq2d}) are always non-negative. Equations (\ref{sdu2sdE}), (\ref{eq11}) and (\ref{eq9}) show that the classical spectral density $({{\rm {\bf E} \cdot {\bf E}}})^{\nu}_{+}$ is directly proportional to the average number of photons per unit volume per unit frequency and does not include quantum zero-point fluctuations. With these definitions, the classical electromagnetic relationship between $({{\rm {\bf E} \cdot {\bf E}}})^{\nu}_{+}$ and $u_+^{\nu}$ in Eq. (\ref{sdu2sdE}) \textit{derives} the apparently approximate Golden Rule relationship between $d_{S \to R}^2$ and $b_{S \to R}$ in Eq. (\ref{eq28}) as an equality for weakly dispersive and approximately transparent media. Thus, Eq. (\ref{eq27}) shows that dipole-strength spectra at positive frequencies are absorptive and dipole-strength spectra at negative frequencies are emissive.

The exact relationship between $A$-coefficient spectral densities and dipole-strength spectra now follows from Eqs. (\ref{eq7}), (\ref{eq14a}) and (\ref{eq28}):
\begin{subequations}
\label{eqDSger}
\begin{flalign}
\begin{split}
\label{eq29}
& a_{S\to R}^{\nu} (-\nu ;P_{S}^{E \eta}) \\
& \quad =
 f_{local}^{2}(\nu) \dfrac{[n(\nu )]^{3}}{\epsilon (\nu)}
\dfrac{8\pi h\nu^{3}}{(cycles)^3c^{3}}
 \dfrac{1}{6\hslash^{2}}
 d_{S\to R}^{2} (-\nu ;P_{S}^{E \eta})\theta (\nu ). 
\end{split}
\end{flalign}
Seven assumptions were used to derive this relationship: $^{\#}1$) Einstein-coefficient spectra (essential); $^{\#}2$) weak coupling and dilute molecules (essential); $^{\#}3$) a linear, homogeneous, and isotropic medium (simplifying); $^{\#}4$) molecular pseudo-isotropy (simplifying); $^{\#}5$) a large cavity (simplifying); $^{\#}6$) electric-dipole approximation (essential); $^{\#}7$) a weakly dispersive and approximately transparent medium (essential). Assumptions $^{\#}1$-5 also appear in the derivation of ref. \onlinecite{RN3964}. Assumptions $^{\#}6$ and $^{\#}7$ are not made there and allow derivation of the ``photon energy $E=h \nu$'' and ``$l$ stimulated +1 spontaneous'' assumptions of ref. \onlinecite{RN3964} through the electric field operator.\cite{RN2514,RN4102,RN5} The Golden Rule time restrictions apply, but energy conservation and a Boltzmann distribution are not assumed. Equation (\ref{eq29}) parallels Eq. (\ref{eq14a}) in that both derivations allow non-equilibrium population distributions $P_{S}^{E \eta}$ within the initial band [see Eq. (\ref{eqAB})] but not coherence. Of course, Eq. (\ref{eq29}) also holds at thermodynamic equilibrium with arguments $V$ and $T$ replacing $P_{S}^{E \eta} = {}^{eq}P_{S}^{E \eta}(V,T)$ in both the $A$-coefficient spectral density and the dipole-strength spectrum.

At this point, we can establish the exact generalized Einstein relation between reverse and forward dipole-strength spectra by invoking their relationship to Einstein $B$-coefficient spectra in Eq. (\ref{eq28}) and relying on the detailed-balance derivation of Eq. (\ref{eq14b}) between Einstein-coefficient spectra for time-reversal invariant systems in ref. \onlinecite{RN3964}.
\begin{flalign}
\begin{split}
\label{eq14d}
& d_{S\to R}^2 (-\nu ,V,T) \\
& \quad \quad =\dfrac{{ }^{eq}\mathcal{P}_{R} (V,T)}{{ }^{eq}\mathcal{P}_{S} 
(V,T)}d_{R\to S}^2 (\nu ,V,T)\exp (-h\nu /k_{\text{B}} T).
\end{split}
\end{flalign}
Here, the dipole-strength spectra have volume and temperature as additional arguments that indicate they are dipole-strength spectra at thermodynamic equilibrium. This derivation of Eq. (\ref{eq14d}) relies on the 7 assumptions used to derive Eq. (\ref{eq29}) above plus two additional assumptions from ref. \onlinecite{RN3964}: detailed balance at equilibrium in a time-reversal invariant system\cite{RN2662,RN3917,RN3915,RN3918,RN3492} and the Planck blackbody radiation spectral density.\cite{RN2744,RN3920,RN3495} The first additional assumption excludes systems in overall rotation and systems in external magnetic fields.\cite{RN3492} Although molecular statistical mechanics was used to motivate dipole-strength spectra (and to provide approximate Golden Rule quantum and statistical formulas for them) in section \ref{secFR}, it was not used in deriving Eq. (\ref{eq14d}). As in the derivation of Eq. (\ref{eq14a}) in ref. \onlinecite{RN3964}, the Boltzmann factor in Eq. (\ref{eq14d}) is imposed by equilibrium with blackbody radiation.

The isotropic assumptions $^{\#}3$ and $^{\#}4$ make the proto-$B$-coefficient independent of the vector directions $\mbox{\boldmath$\varepsilon$}$ and ${\bf k}$ so that a special case of Eq. (\ref{eqPROTObGER}) follows from Eqs. (\ref{eq28}) and (\ref{eq14d}) for isotropic systems with time-reversal invariance. 

We now invoke Boltzmann statistics for the first time in the derivation presented in this paper; most subsequent equations in this section depend on them. Equation (\ref{eq14d}) can be specialized to Boltzmann statistics for ideal behavior (ideal gas, ideal solution, etc.) by using the band population ratio in Eq. (\ref{eqdm}) to generate an ideal system Boltzmann form of the relationship between reverse and forward dipole-strength spectra that parallels Eq. (\ref{eq21}) between $B$-coefficient spectra.
\begin{flalign}
\begin{split}
\label{eq33}
& d_{S\to R}^{2} (-\nu ,V,T) \\
& \quad =d_{R\to S}^{2} 
(\nu ,V,T)\exp [-(h\nu -\Delta \mu_{R\to S}^{\circ} 
(V,T))/k_{\rm{B}} T].
\end{split}
\end{flalign}
For molecules in an ideal-system that obey Boltzmann statistics, Eq. (\ref{eq33}) is an exact relationship for single molecule absorption and emission at thermodynamic equilibrium in a homogeneous, isotropic, and time-reversal invariant system.

The generalized Einstein relation between (reverse) spontaneous emission and the (forward) absorption dipole-strength spectrum holds at equilibrium:
\begin{flalign}
\begin{split}
\label{eq33d}
& a_{S\to R}^{\nu} (-\nu ,V,T) =  \dfrac{[n(\nu )]^{3}}{\epsilon (\nu)}
\dfrac{8\pi h\nu^{3}}{(cycles)^3c^{3}} \theta (\nu )
 \dfrac{f_{local}^{2}(\nu)}{6\hslash^{2}} \\
& \quad \quad \cdot d_{R\to S}^{2} 
(\nu ,V,T)\exp [-(h\nu -\Delta \mu_{R\to S}^{\circ} 
(V,T))/k_{\rm{B}} T].
\end{split}
\end{flalign}
\end{subequations}
Since the standard chemical potential does not depend on frequency, the spontaneous emission lineshape for each molecule is dictated by its absorptive dipole-strength spectrum for positive frequencies and the dielectric properties of the surrounding medium. Macroscopically, the single-molecule relations in Eq. (\ref{eq14d}) - (\ref{eq33d}) can fail for averages over samples in which different molecules have different changes in standard chemical potential.

\subsection{total dipole strength}

The treatment of absorption and emission in total dipole strengths here includes both positive and negative frequencies in the same way as the generalized Einstein $B$-coefficient spectra. The equations assume a kinetic rate description but do not assume that thermal quasi-equilibrium holds until the next paragraph, so volume and temperature are not specified. The total dipole strength is defined as the $t=0$ point of the Fourier transform of the dipole-strength spectrum to the time domain,
\begin{equation}
\label{eq31}
D_{S\to R}^{2} (P_{S}^{E \eta})\equiv \frac{1}{cycles}\int_{-\infty }^{+\infty } {d_{S\to R}^{2} (\nu ; P_{S}^{E \eta} )d\nu } .
\end{equation}
The total dipole strength has $\dim [D_{S\to R}^{2} ]=(charge \cdot length)^{2}$.  Like the Einstein $B$ coefficients defined here, its dimensions are independent of the frequency units. With this definition, the total dipole strength for a transition with a specified direction ($R \to S$) between bands in Eq. (\ref{eq31}) has integration limits from --$\infty$ to +$\infty $. As for the $B$ coefficients, the absorption \textit{vs.} stimulated emission type of the transition becomes mixed if positive and negative frequencies both contribute to the integral. Each total dipole strength $D_{S \to R}^2$ no longer refers solely to the total strength of absorption or stimulated emission, but adds both processes for the molecular transition from band to band in the direction indicated by its subscript. The relationship of such dipole-strength spectra to time-correlation functions is beyond the scope of this paper, but motivates the definition of total dipole strength through a Fourier transform.

If the linewidth is so much narrower than the thermal energy that the lineshape can be approximated as a delta function, integration of quasi-equilibrium Eq. (\ref{eq33}) gives
\begin{flalign}
\begin{split}
&D_{S\to R}^{2}(V,T) \mathop =\limits^{line} 
D_{R\to S}^{2}(V,T)\\
& \quad \quad \quad \quad \quad \quad \cdot \exp [-(h\nu_{R \to S} -\Delta \mu_{R\to S}^{\circ} (V,T))/k_{\rm{B}} T].
\end{split}
\end{flalign}
For line transitions between quantum levels $r$ and $s$ in vacuum, the change in standard partial molecular internal energy becomes the quantum energy-level difference, $\Delta u_{r \to s}^{\circ} = h \nu _{sr}$, and the change in standard partial molecular entropy becomes $\Delta s_{r \to s}^{\circ} = k_{\text{B}} \ln (g_s / g_r)$ where $g_s$ ($g_r$) is the degeneracy of quantum level $s$ ($r$), thus giving $\Delta \mu_{r \to s}^{\circ} = h \nu _{sr} - k_{\text{B}}T \ln (g_s / g_r)$, so that Eq. (\ref{eq33}) further reduces to
\begin{equation}
\label{eq34}
g_{s} D_{s\to r}^{2} \mathop =\limits_{vacuum}^{line} 
g_{r} D_{r\to s}^{2} \mathop =\limits_{vacuum}^{line} \mathsf{S}(r,s).
\end{equation}
The strength for transitions between levels $r$ and $s$ in Eq. (\ref{eq34}) is symmetrical in the initial and final levels, $\mathsf{S}(r,s)=\mathsf{S}(s,r)$, and is the same as the strength $\mathsf{S}(A,B)$ for transitions between levels $A$ and $B$ defined on page 98 of Condon and Shortley\cite{RN725} or the line strength $S_{J'J"}$ in ref. \onlinecite{RN3662}. This states the “principle of reciprocity” for line strengths in vacuum, here derived as a consequence of the generalized Einstein relations. This relationship between the total dipole strengths of absorption and emission lines in vacuum parallels the Einstein relation $g_s B_{s \to r} = g_r B_{r \to s}$. From the perspective of the generalized Einstein relations, the degeneracy factors in the Einstein relations for lines in vacuum are purely entropic factors.

\subsection{transition cross sections}

If sufficiently dilute molecules in a homogeneous sample are isotropically oriented on average, the single-molecule dipole-strength spectra for isotropic and unpolarized light can be used to calculate absorption and stimulated emission from a polarized beam of light. The derivation of the relationship between the transition cross-section spectrum and the dipole-strength spectrum depends on the cycle-averaged relationship between the spectral density of the squared electric field $({{\rm {\bf E} \cdot {\bf E}}})_{+}^{\nu}(\nu )$ and the spectral irradiance $I_{+}^{\nu}(\nu )$ given by Eq. (\ref{sdEEsdI}). This use of frequency-domain quantities in a time-domain rate law requires characteristic timescales much longer than the coherence time of partially coherent beams so that frequency can be specified as a function of time. Based on the definition of dipole-strength spectra, the energy density lost from the beam per unit time through absorption transitions from initial band $R$ to final band $S$ is equal to the product of the molecular number density $N_R$ in the initial band $R$, the energy per photon $h\nu$, and the conditional transition probability per unit time for absorption transitions from $R \to S$ [$d_{R \to S}^2 (\nu , V, T)$ term in Eq. (\ref{eq2d})]. There is also an energy density gain through stimulated emission transitions from $R \to S$ [$d_{R \to S}^2(-\nu , V, T)$ term in Eq. (\ref{eq2d})]. These losses and gains must be summed over all initial and final bands.  Inserting these into the energy conservation equation [Eq. (6.108) of ref. \onlinecite{RN1580}] yields
\begin{flalign}
\begin{split}
\label{eq66}
&\dfrac{\partial I_{+}^{\nu}(z,t;\nu )}{\partial z}+ \dfrac{\partial u_{+}^{\nu}(z,t;\nu )}{\partial t} \\
& \quad \quad =-\sum\limits_{R,S} { 
N_{R} h\nu\dfrac{f_{local}^{2}(\nu) }{6 \hslash^2}
[d_{R\to S}^2 (\nu) - d_{R\to S}^2 (-\nu)]} \\
& \quad \quad \quad \quad \quad \cdot ({{\rm {\bf E} \cdot {\bf E}}})_{+}^{\nu}(z,t;\nu ). 
\end{split}
\end{flalign}
The sum in Eq. (\ref{eq66}) allows the band $R$ to lie below $S$, above $S$, or to be $S$ (an intraband transition). Since dipole-strength spectra are non-negative, the sign of each bracketed difference between absorption and stimulated emission follows the dominant process for that transition at each positive frequency.

In steady state, $\partial u_{+}^{\nu}(z,t;\nu )/\partial t=0$, so that substituting for the positive-frequency spectral density for the square of the electric field on the right-hand side using Eq. (\ref{sdEEsdI}) yields

\begin{flalign}
\begin{split}
\label{eq67}
& \dfrac{\partial I_{+}^{\nu}(z,t;\nu )}{\partial z} \\
& \quad \quad =-\sum\limits_{R,S} { N_{R} h\nu\dfrac{f_{local}^{2}(\nu) }{6 \hslash^2}
[d_{R\to S}^2 (\nu) - d_{R\to S}^2 (-\nu)]} \\ 
& \quad \quad \quad \;\cdot
\dfrac{n(\nu)}{c} \dfrac{1}{\epsilon(\nu)}
I_{+}^{\nu}(z,t;\nu ).
\end{split}
\end{flalign}

\noindent Equation (\ref{eq67}) determines the transition cross-section spectrum for an electric-dipole transition as

\begin{flalign}
\begin{split}
\label{eq68}
&\sigma_{R\to S} (\nu ,V,T) \\
& \quad \quad = \dfrac{h\nu}{c} \dfrac{n(\nu)}{\epsilon(\nu)} \dfrac{1}{6 \hslash^2}\\
& \quad \quad \quad \cdot f_{local}^{2}(\nu) [d_{R\to S}^2 (\nu ,V,T) - d_{R\to S}^2(-\nu ,V,T)],
\end{split}
\end{flalign}

\noindent so that Eq. (\ref{eq67}) becomes
\begin{equation}
\label{eq16}
\dfrac{\partial I_{+}^{\nu}(z,t;\nu )}{\partial z}=-\sum\limits_{R,S} {N_{R} \sigma_{R\to S} (\nu ,V,T)} I_{+}^{\nu}(z,t;\nu ).
\end{equation}
If the number densities are independent of $z$, this integrates to
\begin{flalign}
\begin{split}
\label{eqBL}
I_{+}^{\nu}(z,t;\nu ) = & I_{+}^{\nu}(z=0,t;\nu ) \\
& \cdot \exp [-\sum\limits_{R,S} {N_{R} \sigma_{R\to S} (\nu ,V,T)}z ].
\end{split}
\end{flalign}

Since the refractive index and dielectric constant are even functions of frequency,\cite{RN2558} the above transition cross sections are each an even function of frequency over the entire real frequency axis with 
$\sigma_{R\to S} (\nu ,V,T)=\sigma_{R\to S} (-\nu ,V,T)$.
The transition cross sections in Eq. (\ref{eq68}) have a definite direction from band to band specified by the subscript arrow, but need not have a definite absorption or stimulated emission type. Each transition cross section here is positive for frequencies at which absorption dominates, negative for frequencies at which stimulated emission dominates, and vanishes as $\nu ^2$ (or a higher even power) towards zero frequency. If only two bands are populated and one lies entirely above the other, Eq. (\ref{eq16}) reduces to have the same effect as Eq. (51) in Sec. 7.5 of ref. \onlinecite{RN3162}, but it has a different form because of two compensating sign differences: the stimulated emission cross section is negative in Eq. (\ref{eq68}) here \textit{vs.} positive in ref. \onlinecite{RN3162} and the sign in front of stimulated emission cross sections remains negative in Eq. (\ref{eq16}) \textit{vs.} specially reversed in ref. \onlinecite{RN3162}.

In contrast to the parallel relationship between the transition cross-section and $B$-coefficient spectra in Eq. [18] of ref. \onlinecite{RN3964}, the frequency derivative of the refractive index from the energy transport velocity does not appear in Eq. (\ref{eq68}). Similarly, the derivative of the refractive index does not appear in the relationship between the spontaneous emission spectral density $a_{S \to R}^{\nu}(-\nu)$ and stimulated emission dipole-strength spectrum $d_{S \to R}^2 (-\nu)$ in Eq. (\ref{eq29}). Equations (\ref{eq29}) and (\ref{eq68}) suggest that the dispersive factors in the $B$-coefficient spectra are an unfortunate consequence of defining the Einstein $B$ coefficients in terms of the electromagnetic energy density.

If the medium is non-magnetic,
$\epsilon (\nu )=[n(\nu )]^{2}$, and one obtains
\begin{equation}
\label{sigd2NMeq}
 \sigma_{S\to R} (\nu )
 =\left[ \dfrac{f_{local}^{2}(\nu)}{n (\nu )} \right]
 \dfrac{h\nu}{c}
 \dfrac{1}{6\hslash^{2}} 
[d_{S\to R}^{2} (\nu ) - d_{S\to R}^{2} (-\nu )] \\ 
\end{equation}
and
\begin{equation}
\label{afromd2nm}
\begin{array}{c}
 a_{S\to R}^{\nu} (-\nu )=[f_{local}^{2}(\nu) n(\nu )]
 \dfrac{8\pi h\nu^{3}}{(cycles)^3c^{3}}\\
\quad \quad \quad \cdot \dfrac{1}{6\hslash^{2}}
 d_{S\to R}^{2} (-\nu )\theta (\nu ). \\  
 \end{array}
\end{equation}
The prefactors in square brackets capture the entire medium dependence. This refractive-index dependence of the spontaneous emission spectral density reduces to that for line spectra in refs. \onlinecite{RN4102,RN5,RN3993}; ref. \onlinecite{RN3992} also provides a dielectric continuum formula for the local field factor. Glauber and Lewenstein \cite{RN3994, RN1, RN3996} found the same refractive-index dependence with a different (``empty-cavity'' - see ref. \onlinecite{RN175}) treatment of the local field.
\subsection{narrow spectra}
The relationship between the transition cross-section spectra and dipole-strength spectra simplifies if the linewidth (variance $\Delta^2$) is sufficiently narrow with respect to both the thermal energy $k_{\rm B} T$ and the center frequency $\nu_{R \to S}$; ($h \Delta ^2 /k_{\rm B} T + \Delta/2 ) \ll \nu_{R \to S}$. With the above restrictions, one dipole-strength spectrum is confined to positive frequencies so that it represents absorption and its reverse is confined to negative frequencies so that it represents stimulated emission. In consequence, if the frequency is taken to be positive, only one of the two dipole strength terms on the right hand-side of Eq. (\ref{sigd2NMeq}) is appreciable and we have the forward absorption cross-section spectrum
\begin{subequations}
\label{sigd2POSeq}
\begin{equation}
\label{sigd2POSeqa}
 \sigma_{R\to S} (\nu ) \theta(\nu)
 \approx \left[ \dfrac{f_{local}^{2}(\nu)}{n (\nu )} \right]
 \dfrac{h\nu}{c}
 \dfrac{1}{6\hslash^{2}} 
d_{R\to S}^{2} (\nu ) \theta(\nu),\\ 
\end{equation}
and the reverse stimulated emission cross-section spectrum
\begin{equation}
\label{sigd2POSeqb}
 \sigma_{S \to R} (\nu ) \theta(\nu)
 \approx - \left[ \dfrac{f_{local}^{2}(\nu)}{n (\nu )} \right]
 \dfrac{h\nu}{c}
 \dfrac{1}{6\hslash^{2}} 
d_{S\to R}^{2} (-\nu ) \theta(\nu).\\ 
\end{equation}
\end{subequations}
Again, the prefactors in square brackets capture the entire solvent dependence. In this approximation, Eq. (\ref{eq33}) shows that the absorption and stimulated emission cross sections directly obey a generalized Einstein relation
\begin{flalign}
\begin{split}
\label{eq33sigma}
& \sigma_{S\to R} (\nu ,V,T) \theta(\nu) \approx - \sigma_{R\to S} (\nu ,V,T) \\
& \quad \quad \quad \quad \cdot \exp [-(h\nu -\Delta \mu_{R\to S}^{\circ} (V,T))/k_{\rm{B}} T] \theta(\nu).
\end{split}
\end{flalign}
Accounting for sign conventions, this is an isotropic version of the relationship found by McCumber,\cite{RN2267} clarified by use of a standard chemical potential.\cite{RN2935,RN3964} (With the delta-function approximation to the Golden Rule, Eq. (\ref{eqPROTObGER}) would establish this as the single-mode relationship found by McCumber.) In this approximation, Eqs. (\ref{afromd2nm}), (\ref{sigd2POSeqb}) and (\ref{eq33sigma}) show that the equilibrium spectral density for spontaneous emission is related to the absorption cross-section spectrum by
\begin{flalign}
\begin{split}
\label{afromsigmaPOS}
& a_{S\to R}^{\nu} (-\nu ,V,T) \approx \dfrac{8\pi \nu^2}{(cycles)^3}
 \left( \dfrac{ n(\nu )}{c} \right)^2
 \sigma_{R\to S} (\nu ,V,T) \\ 
&\quad \quad \quad \quad \quad \quad \cdot \exp [-(h\nu -\Delta \mu_{R\to S}^{\circ}(V,T))/k_{\rm{B}} T]\theta (\nu ).
\end{split}
\end{flalign}
Integrating both sides over frequency gives a generalization of the Strickler-Berg formula for the radiative rate [Eq. (21) of ref. \onlinecite{RN3262}] that does not make the Born-Oppenheimer approximation (ref. \onlinecite{RN3262} retained only one Born-Oppenheimer basis state) or the Condon approximation\cite{RN723} that the electronic transition dipole is independent of the vibrational coordinates. The change in standard chemical potential in Eq. (\ref{afromsigmaPOS}) can alter the relationship between absorption cross section and radiative lifetime (see subsection \ref{scp} below) but is not present in the Strickler-Berg formula. This generalization also allows for dispersion of the refractive index, but assumes thermal quasi-equilibrium, which is not required by the  Strickler-Berg formula.

\section{Discussion}

The quantized treatment of electric-dipole transitions in dispersive media presented here has provided new generalized Einstein relations [Eq.(\ref{eqDSger})] between equilibrium dipole-strength spectra for absorption and emission at the single molecule level. In comparison to the derivation in ref. \onlinecite{RN3964}, the electric-dipole approximation and quantized electric field operator are used here to derive the ``photon energy $E=h \nu$'' and ``$l$ stimulated +1 spontaneous'' premises assumed there. The consequent relationships between dipole-strength spectra are rigorously applicable to single-photon electric-dipole transitions of single molecules in dispersive media that are approximately transparent. Up to Eq. (\ref{eqDSger}) and from Eq. (\ref{eq66}) to (\ref{afromd2nm}), spectra can be narrow or broad with respect to the thermal energy, spectra can be narrow or broad with respect to the photon energy, and photon energies can be small or large compared to the thermal energy as in ref. \onlinecite{RN3964}. The relationships between the dipole-strength spectra, transition cross-section spectra, and spontaneous emission spectral densities depend on the refractive index, dielectric constant, and local field factor, but do not depend on the frequency derivative of the refractive index. Since the electromagnetic energy density in dispersive media has been a complicated subject (see refs. \onlinecite{RN476,RN477,RN4102,RN5,RN832,RN1580,RN1985,RN3987,RN3989,RN3990} and Appendix A) these results from the field quantization\cite{RN2514,RN4102,RN5} are reassuring. They suggest that, for electric-dipole transitions, the single-molecule dipole-strength spectra will be more practically useful than Einstein $B$-coefficient spectra in dispersive media .

In practice, the restriction that the dispersive medium be ``approximately transparent'' requires only that absorptive losses be small over a distance of one wavelength, not that the absorbance be small over the sample length. For example, near 500 nm wavelength, a solution that transmits 10\% of the incident energy through a distance of 100 $\mu$m has a smaller absorptive loss of $\sim$1\% over a distance of one wavelength and the neglected imaginary part of its complex-valued refractive index is $\sim \! 9 \! \times \! 10^{-4}$.\cite{RN90} Use of the electromagnetic density of modes and molecular properties within an infinite medium requires that any cavity be ``large'' compared to the wavelength of light and weak coupling requires that any cavity be ``bad''.\cite{RN2315} 

For homogeneously broadened absorption and stimulated emission spectra, solutions are ``sufficiently dilute'' if they practically eliminate chemical (e.g. aggregation), quantum statistical, and electromagnetic (e.g. transition dipole) couplings that shift levels or alter transition dipole moments.  Here, the relevant timescale for distinguishing between homogeneous and inhomogeneous broadening is the excited state lifetime. In these circumstances, the macroscopic absorption spectra in Eq. (\ref{eqBL}) are connected to transition cross sections and dipole-strength spectra through Eq. (\ref{eq68}) or (\ref{sigd2NMeq}).

The equilibrium spontaneous emission connection to absorption in Eq. (\ref{eq33d}) is also approximately applicable to non-equilibrium photoluminescence when thermal quasi-equilibrium within the upper band is obtained on a timescale much faster than the upper band lifetime (thus making the photoluminescence lineshape independent of the excitation wavelength\cite{RN1736,RN3239}). In addition to the ``sufficiently dilute'' criteria for stimulated transitions, practical application of the single-molecule results to luminescence requires concentrations far below\cite{RN2971} the ``critical concentration'' given by Eq. (20) of ref. \onlinecite{RN1111} to limit energy transfer during the excited state lifetime. The derivation of the generalized Einstein relations does not rely on the adiabatic or Born-Oppenheimer approximation, but the condition for applicability to non-equilibrium photoluminescence requires an analogous separation of fast intraband \textit{vs.} slow interband relaxation timescales. For example, groups of molecular electronic states that are rapidly coupled together by non-adiabatic processes should be lumped together into a single band.

The rigorous derivation of Eq. (\ref{eq14d}) between dipole-strength spectra for stimulated forward and reverse transitions (e.g. absorption and stimulated emission) also depends on time-reversal invariance, and thus excludes rotating systems and systems in static external magnetic fields. However, if the \textit{ad hoc} hypothesis of Eq. (\ref{rGER}) is correct, then this restriction to time-reversal invariant systems is not essential and Eq. (\ref{eqPROTObGER}) would be applicable to steady-state Boltzmann distributions for systems lacking time-reversal invariance.  

\subsection{\label{scp} the standard chemical potential}

From an experimental point of view, spectroscopic measurement of the change in standard chemical potential requires determination of the forward and reverse $B$-coefficient spectra [Eq. (\ref{eq14b})] or dipole-strength spectra [Eq. (\ref{eq14d})] at one frequency. This method of determination does not imply that the change in standard chemical potential between two bands has a physical dependence on the transition moments between them. Each band's standard chemical potential from Eq. (\ref{eqSCP}) depends only on its level structure, the temperature, and the volume. The analogy to the Einstein relations for line spectra in vacuum is that ratios of $B$-coefficients or total dipole strengths [Eq. (\ref{eq34})] determine the ratio of degeneracies, which are solely properties of the levels. Just as the degeneracy ratio plays a dominant role in calculating the radiative lifetime from absorption line spectra in vacuum [see ref. \onlinecite{RN1457}], the change in standard chemical potential played a dominant role in calculating the radiative lifetime from quantum dot absorption spectra [see Eqs. (22) and (S14) plus the last two paragraphs of ``Tests of the generalized Einstein relations'' in ref. \onlinecite{RN2971}].

\subsection{the local field}
For comparison to experiment, any theory of refractive effects on transition strengths and lifetimes must include local field enhancements. Beyond the refractive effects for line spectra treated by Nienhuis and Alkemade,\cite{RN2514} and by Milonni\cite{RN4102,RN5} our results give refractive index and local field effects on broadband electric-dipole absorption and emission in dispersive media. The local field effect inside a medium can be treated either as enhancement of the field, as enhancement of the transition dipole, or as a combination (see \S 9 of ref. \onlinecite{RN422}). Consistent with any of these approaches, the local field factor and dipole strength spectra always appear together as a product in expressions for the spontaneous emission spectral density and stimulated transition cross-section spectra. The results obtained here can be verified against the expressions for refractive effects inside a nondispersive medium that were elegantly tested by Lamouche \textit{et al.}\cite{RN1979} 

Local fields effects are often just as significant as refractive index and dielectric constant effects for electric-dipole transitions in dielectrics,\cite{RN2902} but can be obscured by specific solvent-solute interactions that are not contained within electromagnetic theory.\cite{RN1979} The refractive index and dielectric constant effects in Eqs. (\ref{sigd2NMeq}) and (\ref{afromd2nm}) are in agreement with experimental results on solvent mixtures containing single dye molecules inside inverse micelles that are much smaller than the wavelength of light.\cite{RN1979} These beautiful studies conceptually include the entire interior of the inverse micelle in calculating the electric-dipole transition moment and calculate the solvent mixture dependent local field factor for the inverse micelle from macroscopic electromagnetic theory. This use of a local-field factor is possible because the micelles (radius $\sim$ 1.5 nm) are much smaller than the wavelengths of light involved ($\sim$ 600 nm).

Since Lorentz, the treatment of local-field effects has often incorporated radiative level shifts from the molecular environment.\cite{RN2143} Milonni has used the Lorentz-Lorenz local field to treat radiative level shifts for line spectra in a dispersive dielectric.\cite{RN4102} For a non-polar molecule, these level shifts can be considered to arise from van der Waals interactions with polarizable non-polar molecules in their environment,\cite{RN4102} as for the ``general red-shift'' of spectra in condensed phases.\cite{RN284,RN3805} Because the radiative shift of a level depends on all of the states that are radiatively coupled to it\cite{RN4102} (including many that may have specific resonances with the environment) and because of the specific interactions mentioned in the paragraph above, these level shifts should be included in the molecule + environment energies here.

\subsection{\label{dv} domain of validity}
In contrast to Eq. (\ref{eq2a}), which is restricted to infinite, homogeneous, and isotropic media, Eq. (\ref{eq2d}) is valid for electric-dipole transitions in isotropic media even if the spectral density for the square of the field varies with spatial position (e.g. standing waves).  For electric-dipole transitions in infinite, homogeneous, and isotropic media, Eq. (\ref{eq2b}) for the conditional transition probability per unit time for a molecule in band $S$ to make a spontaneous single-photon transition to band $R$ holds with $a$ from Eq. (\ref{eq29}). However, if the spectral density for the square of the electric field in Eq. (\ref{eq2d}) is spatially dependent, then factors in Eq. (\ref{eq29}) arising from the spectral density of modes must be reformulated in terms of the spatially-dependent spectral density for fluctuations of the square of the electric field, which is beyond the scope of this paper.
 
Following Dirac, the Golden Rule was used to treat single-photon transitions in a quantized field, which is essential for relating spontaneous and stimulated emission. If the Golden Rule's short-time criterion can be met, then the integrated conditional transition probability per unit time that Dirac used to find the Einstein $A$ and $B$-coefficients does not change with increasing time and his derivation is valid for the coefficients. However, the restricted time interval in the derivation of the Golden Rule expression for spectra do not allow it to provide a comprehensive derivation of the generalized Einstein relations between spectra. Let us consider what would happen in Eq. (\ref{eqPROTObRATIO}). We have seen that the delta function approximation would lead to proto-$B$-coefficient spectra that would obey the generalized Einstein relation in Eq. (\ref{eqPROTObGER}), that these are proportional to the Einstein $B$-coefficient spectra in homogeneous and isotropic media, and that the $B$-coefficient spectra must obey the generalized Einstein relation in Eq. (\ref{eq14b}) to satisfy the detailed-balance condition in Eq. (\ref{eq4}) with Planck blackbody radiation. Replacing the infinite time delta function limit of $r_t$ with any symmetric non-negative function $r_t(\Delta E)$ that has a non-zero width would convolve both spectra, broadening them without increasing the linewidth-dependent Stokes' shift between them. In consequence, the symmetric functions $r_t$ from the Golden Rule cannot satisfy detailed balance with Planck blackbody radiation, and neither can the Lorentzian from extending the Golden Rule treatment to longer times in complement $\rm{D_{XIII}}$ of ref. \onlinecite{RN711} (a Lorentzian is derived from the Weisskopf-Wigner approximation\cite{RN3900,RN2315} that the spectral density of radiation modes is flat, which is why it fails).

The Golden Rule approximates quantum dynamics to first order in the perturbation for a single absorption or emission event in an isolated system, while the generalized Einstein relations are concerned with the exact long-time average spectra for rates at which many events occur in thermodynamic equilibrium. This key difference does not justify the approximate delta function limit of the Golden Rule, which enforces exact energy conservation for the system: Planck blackbody radiation requires a definite temperature, and a fixed temperature for the system requires fluctuations in the system's energy through interaction with the thermal reservoir;\cite{RN2661,RN2286,RN594} with thermal fluctuations in the system's energy, the time-energy uncertainty principle\cite{RN3980} requires broadening of the system's energy levels. A pair of functions $r_{RE\eta,SE'\eta'}$ and $r_{SE'\eta',RE\eta}$ for the forward and reverse transitions that are related by Eq. (\ref{rGER}) provide a way to exactly satisfy detailed balance without doing obvious essential violence to statistical or quantum mechanics. A theory for these functions might be sought in the thermal-reservoir induced decoherence\cite{RN541,RN3241,RN2415} between $|SE'\eta '\rangle$ and $|RE\eta \rangle$, which is neglected (or effectively swept into the density of coupled final states) in the Golden Rule, but it would not capture purely radiative lifetime broadening of an initial upper state.\cite{RN4109}

In some sense, the purpose of this Golden Rule treatment is to understand as much of the forward and reverse transition decoherence broadening processes as possible by bringing key degrees of freedom into the system. Comparing the Golden Rule expressions obtained here for the forward and reverse Einstein $B$-coefficient spectra in the numerator and denominator of Eq. (\ref{eqPROTObRATIO}) and ignoring the pair of functions $r$, the difference arises from thermally averaging over the initial band \textit{vs.} summing over the final band; the magnitude-squared transition matrix elements, the densities of states, and the radiative effects from the environment are all the same in both directions. These aspects appear in the more-restricted formulas of McCumber\cite{RN2267} and, along with a relationship between forward and reverse functions $r$, are the statistical and quantum mechanical keys to the generalized Einstein relations for time-reversal invariant systems.

\section{CONCLUSIONS}

We have used the Golden Rule and a Boltzmann distribution in a quantum and statistical mechanical approach to equilibrium relationships between absorption and emission spectra in the electric-dipole approximation.  Generalized Einstein relations between dipole-strength spectra emerge from thermal averaging within the initial band, the Hermitian character of quantum matrix elements, and summation within the final band. This rate theory is limited to spectra broadened by dynamics within the initial and final bands, but provides a route to construct spectra that obey the generalized Einstein relations for quantum models. The treatment of electric-dipole transitions here includes the effects on transition strength and lineshape from the dielectric constant, refractive index, and dispersion in the medium surrounding the absorber/emitter, including local field effects. The new single-molecule generalized Einstein relations between dipole-strength spectra, like those between homogeneous absorption and stimulated emission cross sections for sufficiently narrow transitions at high frequency,\cite{RN2267} depend only on the refractive index at each frequency, and do not depend on the frequency derivatives of the refractive index. The calculated refractive index  dependence of the cross sections and radiative lifetime agree with experiment for line spectra, and the local field dependence agrees with that established for sufficiently narrow spectra.

Although the relationships apply rigorously to thermodynamic equilibrium emission, they are expected to apply with high accuracy to the photoluminescence of homogeneous samples that relax to thermal quasi-equilibrium within the excited band on a timescale very much faster than the excited band's population lifetime. The generalized Einstein relations between forward and reverse dipole-strength spectra enable spectroscopic determination of the difference in standard chemical potential between bands. The standard chemical potential of each band is dictated by the quantum statistical properties of that band in the same way that the standard chemical potential of a molecule is dictated by its properties. The connections obtained here between standard chemical potentials, transition cross sections, lineshapes, radiative rates, and the refractive index open up a route to spectroscopic thermodynamics. 

\begin{acknowledgments}
We thank Casey Hynes for discussions of the Golden Rule and Peter Milonni for confirming the corrections in footnote \onlinecite{RN5}. We thank Aman Agrawal, Jérémie Léonard, David McCamant, and Jennifer Sormberger for helpful comments and questions. This material is based upon work supported by the National Science Foundation under Award No. CHE-2155010.
\end{acknowledgments}

Conflict of Interest Statement – The authors have no conflicts to disclose.

Author Contributions – D.M.J. designed research; J.R. and D.M.J. developed the Golden Rule treatment, dipole-strength spectra, transition cross sections, the standard chemical potential, and Appendices C \& D; D.M.J. developed the other treatments, J.R. and D.M.J. wrote the initial draft with rewriting by both before development of ref. \onlinecite{RN3964} and rewriting mainly by D.M.J. afterwards.

Data Availability - Data sharing is not applicable to this article as no new data were created or analyzed in this study.

\renewcommand{\theequation}{A-\arabic{equation}}
\setcounter{equation}{0}
\section*{Appendix A}

This appendix relates the positive-frequency spectral density of electromagnetic energy per unit volume in the Planck blackbody radiation law and the Einstein relations to the frequency-domain fields arising from complex exponential Fourier transforms, which extend over positive and negative frequencies. A reader who is bothered by negative frequencies should consider that positive \textit{vs.} negative rotational velocities quantify the clockwise \textit{vs.} counter-clockwise rotation of a wheel. With complex exponential Fourier transforms, both positive and negative frequencies are needed to form the complete orthogonal basis of sines and cosines. 

We start by determining the spectrum and the spectral density of the irradiance. We start with irradiance because the frequency domain fields for partially coherent light have some indeterminate 2D random walk character\cite{RN3986,RN3985} and the electromagnetic energy density in dispersive media is complicated.\cite{RN2514,RN476,RN477,RN3987,RN832,RN1985,RN1580,RN3989,RN3990,RN4102,RN4103} At position $\rm{\bf{x}}$ and time $t$, the classical time-domain electromagnetic energy flux is given by the Poynting vector
\begin{equation}
\label{tdS}
{\rm {\bf S}}({\rm {\bf x}},t)
= c
 {\rm {\bf E}}({\rm {\bf x}},t)\times{\rm {\bf H}}({\rm {\bf x}},t),
\end{equation}
\noindent where $c$ is the speed of light in vacuum, ${\rm {\bf E}}({\rm {\bf x}},t)$ is the electric field vector, and ${\rm {\bf H}}({\rm {\bf x}},t)$ is the magnetic intensity vector. Eq. (\ref{tdS}) is written in Heaviside-Lorentz units,\cite{RN1580} where $\epsilon $ and $\mu$ are dimensionless and $\dim [{\rm {\bf E}}({\rm {\bf x}},t)]$ = $\dim [{\rm {\bf H}}({\rm {\bf x}},t)]$ = $mass^{1/2} /(length^{1/2} \cdot time)$ so that $\dim [{\rm {\bf S}}({\rm {\bf x}},t)]$ = $mass / time^3$.  For a plane wave, the irradiance is the scalar absolute magnitude of the energy flux vector
\begin{equation}
\label{tdI}
I({\rm {\bf x}},t) = 
|{\rm {\bf S}}({\rm {\bf x}},t)|.
\end{equation}
\noindent We are interested in the stationary time-averaged irradiance for partially coherent light sources, which we will access through the time-averaged Poynting vector
\begin{equation}
\label{taS}
\langle{ {\rm {\bf S}}({\rm {\bf x}}) } \rangle _t = 
\lim\limits_{T \to \infty} \dfrac{1}{T}
\int_{-T/2}^{+T/2}{\rm {\bf S}}({\rm {\bf x}},t)dt,
\end{equation}
\noindent by extending it to become a time-averaged time-domain correlation function
\begin{align}
\begin{split}
\label{tcS}
& \langle{ {\rm {\bf S}}_{\rm{\bf EH}}({\rm {\bf x}},\tau) } \rangle _t \\
& \quad \quad = 
\lim\limits_{T \to \infty} \dfrac{1}{T}
\int_{-T/2}^{+T/2}c{\rm {\bf E}}({\rm {\bf x}},t)\times{\rm {\bf H}}({\rm {\bf x}},t + \tau)dt.
\end{split}
\end{align}
The subscript $\bf{EH}$ indicates the correlation function of $\bf{E}$ and $\bf{H}$. The frequency spectrum of this time-averaged time-domain correlation function is obtained through its inverse Fourier transform as
\begin{subequations}
\label{cS}
\begin{equation}
\label{fcS}
\langle{ {\rm {\bf \hat{S}}}_{\rm{\bf EH}}({\rm {\bf x}},\nu) } \rangle _t = 
\int_{-\infty}^{+\infty} 
\langle{ {\rm {\bf S}}_{\rm{\bf EH}}({\rm {\bf x}},\tau) } \rangle _t 
\exp(i 2 \pi _c \nu \tau)d \tau.
\end{equation}
for cyclic frequency $\nu$, or as
\begin{equation}
\label{acS}
\langle{ {\rm {\bf \hat{S}}}_{\rm{\bf EH}}({\rm {\bf x}},\omega) } \rangle _t = 
\int_{-\infty}^{+\infty} 
\langle{ {\rm {\bf S}}_{\rm{\bf EH}}({\rm {\bf x}},\tau) } \rangle _t 
\exp(i \omega \tau)d \tau,
\end{equation}
\end{subequations}
for angular frequency $\omega = 2 \pi _c \nu$. Here and in Appendix D, we use $2 \pi _c$ as a compact shorthand for $2 \pi /(cycles)$.\cite{RN3931} Note that with this choice of the inverse Fourier transforms, the spectrum of the correlation function is independent of the frequency dimensions
\begin{equation}
\label{fcac}
\langle{ {\rm {\bf \hat{S}}}_{\rm{\bf EH}}({\rm {\bf x}},\nu) } \rangle _t
=
\langle{ {\rm {\bf \hat{S}}}_{\rm{\bf EH}}({\rm {\bf x}},\omega = 2 \pi _c \nu) } \rangle _t,
\end{equation}
and has $\dim[\langle{ {\rm {\bf \hat{S}}}_{\rm{\bf EH}}({\rm {\bf x}},\nu) } \rangle _t]$ = $mass/time^{2}$. The complex-exponential inverse Fourier transforms in Eq. (\ref{cS}) spread the spectrum of the real-valued correlation function \textit{symmetrically} over both positive and negative frequencies.

Now, the Fourier transform of the time-averaged frequency-domain correlation function spectrum is the time-averaged time-domain correlation function of the lag $\tau$,
\begin{align}
\begin{split}
\label{FTcS}
& \langle{ {\rm {\bf S}}_{\rm{\bf EH}}({\rm {\bf x}},\tau) } \rangle _t \\
& \quad \quad = 
\dfrac{1}{cycles}\int_{-\infty}^{+\infty} 
\langle{ {\rm {\bf \hat{S}}}_{\rm{\bf EH}}({\rm {\bf x}},\nu) } \rangle _t 
\exp(-i 2 \pi _c \nu \tau)d \nu, \\
& \quad \quad = 
\dfrac{1}{2 \pi}\int_{-\infty}^{+\infty} 
\langle{ {\rm {\bf \hat{S}}}_{\rm{\bf EH}}({\rm {\bf x}},\omega) } \rangle _t 
\exp(-i \omega \tau)d \omega.
\end{split}
\end{align}
From Eq. (\ref{FTcS}), the time-averaged Poynting vector can be written in terms of the zero-lag Fourier transform of the spectrum of its time-averaged time-domain correlation function
\begin{align}
\begin{split}
\label{Sfromf}
\langle{ {\rm {\bf S}}({\rm {\bf x}}) } \rangle _t
&=
\langle{ {\rm {\bf S}}_{\rm{\bf EH}}({\rm {\bf x}},\tau = 0) } \rangle _t \\
&= \dfrac{1}{cycles}\int_{-\infty}^{+\infty} 
\langle{ {\rm {\bf \hat{S}}}_{\rm{\bf EH}}({\rm {\bf x}},\nu) } \rangle _t d \nu,
\end{split}
\end{align}
so the spectral density for the Poynting vector is
\begin{equation}
\label{sdS}
{\rm {\bf \hat{S}}}^{\nu}({\rm{\bf x}}, \nu) \equiv 
\dfrac{\partial}{\partial \nu}
\langle{{\rm {\bf S}}({\rm {\bf x}}) } \rangle _t = 
\dfrac{1}{cycles} 
\langle{ {\rm {\bf \hat{S}}}_{\rm{\bf EH}}({\rm {\bf x}},\nu) } \rangle _t.
\end{equation}
The spectral density for the Poynting vector in Eq. (\ref{sdS}) differs from the spectrum of its correlation function in Eq. (\ref{fcS}) by a dimensional factor, which is only $1/cycles$ for cyclic frequencies but can take on a numerical value or change the shape of the spectral density (see examples in the next two paragraphs below).

Now, from Eq. (\ref{tdI}), the spectral density for the irradiance is the absolute value of the spectral density for the Poynting vector,
\begin{equation}
\label{sdI}
I^{\nu}({\rm{\bf x}}, \nu) =
|{\rm {\bf \hat{S}}}^{\nu}({\rm{\bf x}}, \nu)|
= 
|\dfrac{1}{cycles} 
\langle{ {\rm {\bf \hat{S}}}_{\rm{\bf EH}}({\rm {\bf x}},\nu) } \rangle _t|,
\end{equation}
which has $\dim[I^{\nu}({\rm{\bf x}}, \nu)]= mass /(time^2 \cdot cycles)$.
A change of variables transforms the spectral density. For example, as a function of angular frequency, the spectral density for the irradiance is
\begin{equation}
\label{sdIomega}
I^{\omega}({\rm{\bf x}}, \omega)
\equiv 
|\dfrac{\partial}{\partial \omega}
\langle{{\rm {\bf S}}({\rm {\bf x}}) } \rangle _t |= 
\dfrac{1}{2 \pi} 
|\langle{ {\rm {\bf \hat{S}}}_{\rm{\bf EH}}({\rm {\bf x}},\omega) } \rangle _t|.
\end{equation}
These different spectral densities for the irradiance are connected by the change of variables theorem:
\begin{equation}
\label{sdcv}
I^{\nu}({\rm{\bf x}}, \nu)
=I^{\omega}({\rm{\bf x}}, \omega = 2 \pi _c \nu) 
\dfrac{\partial \omega }{\partial \nu}.
\end{equation}

In many applications, the spectral density for the irradiance is considered to be a non-zero function only for positive frequencies, so the positive-frequency spectral density for the irradiance is \textit{twice} the spectral density from the complex-exponential inverse Fourier transforms for positive frequencies (see pp. 179-180 of ref. \onlinecite{RN447})
\begin{equation}
\label{pfsdI}
I^{\nu}_{+}({\rm{\bf x}}, \nu) =
2 I^{\nu}({\rm{\bf x}}, \nu) \theta (\nu),
\end{equation}
where $\theta (\nu )$ is the Heaviside unit step function.\cite{RN447} The positive-frequency spectral density for the irradiance integrates according to
\begin{equation}
\label{Ipfsd}
I({\rm {\bf x}}) = 
\int_{0}^{+\infty} 
I^{\nu}_{+}({\rm {\bf x}},\nu)d \nu.
\end{equation}
The positive-frequency spectral density for the irradiance can be transformed to and from other dependent variables that are conventionally considered to be positive, such as the photon energy $E$ or the wavelength $\lambda$ (dim$[\lambda]=length/cycle$):
\begin{align}
\label{pfsdcv}
\begin{split}
 I^{\nu}_{+}({\rm{\bf x}}, \nu)
  &=I^{E}_{+}({\rm{\bf x}}, E = h \nu) \dfrac{\partial E }{\partial \nu} \\
  &=I^{\lambda}_{+}\left({\rm{\bf x}},
  \lambda = \dfrac {c}{\nu n(\nu)} \right)
  \! \! \left(- \dfrac{\partial \lambda }{\partial \nu} \right).
\end{split}
\end{align}
A negative sign was put into the last change of variables factor in order to reverse the limits in the integral such that the upper limit is the longer wavelength.  These spectral densities directly integrate in the same way as Eq. (\ref{Ipfsd}), for example, the spectral density of the irradiance as a function of wavelength integrates to give the total irradiance
\begin{equation}
\label{sdIE}
I({\rm {\bf x}}) = \int_0^{\infty}I^{\lambda}_{+}({\rm{\bf x}}, \lambda) d\lambda.
\end{equation}
An ideal energy- or wavelength-dispersive spectrometer measures the spectral density $I^{E}_{+}({\rm{\bf x}}, E)$ or $I^{\lambda}_{+}({\rm{\bf x}}, \lambda)$, respectively. The change of variables factor $(\partial \lambda / \partial \nu)$ alters the shape of the spectral density as a function of wavelength \textit{vs.} the spectral density as a function of frequency.

The classical electromagnetic energy in dispersive media is a complicated topic.\cite{RN2514,RN4102,RN476,RN477,RN3987,RN1985,RN3989,RN3990} We will access it using the Poynting vector for a monochromatic wave plus superposition. In transparent isotropic media, Maxwell's equations have traveling monochromatic plane wave solutions with
\begin{equation}
\label{Etrav}
{\rm {\bf E}} ({\bf{x}} , t)
={\rm{\bf E}}_0
\cos( 2 \pi _c \nu t - {\bf{k}} \cdot {\bf{x}} + {\phi}_0 )
\end{equation}
and
\begin{equation}
\label{Htrav}
{\rm {\bf H}} ({\bf{x}} , t)
={\rm{\bf H}}_0
\cos( 2 \pi _c \nu t - {\bf{k}} \cdot {\bf{x}} + {\phi}_0 )
\end{equation}
where
\begin{equation}
\label{HfromEvector}
{\rm {\bf H}}_{0} 
= \dfrac {c}
{2\pi _c \nu \mu (\nu)} \textbf{k} \times
{\rm {\bf E}}_{0}
\end{equation}
\noindent and
\begin{equation}
\label{k}
\textbf{k} \cdot \textbf{k}
=(2\pi _c \nu )^{2} \epsilon (\nu ) \mu (\nu )/c^{2},
\end{equation}
where \textbf{k} is a real-valued wavevector and $\phi _0$ is the time-domain phase.  Substituting these relationships in Eq. (\ref{tdS}) gives the Poynting vector
\begin{align}
\label{tS}
\begin{split}
{\rm {\bf S}}({\rm {\bf x}},t) = &
\; c \left[ \dfrac{\epsilon (\nu)}{\mu (\nu)} \right]^{1/2}
\! \! \! ({\rm {\bf E}}_{0} \cdot {\rm {\bf E}}_{0})
\dfrac{\bf{k}}{|\bf{k}|} \\
& \cdot \cos ^2 (2 \pi _c \nu t - {\bf{k}} \cdot {\bf{x}} + {\phi}_0).
\end{split}
\end{align}

Since Maxwell's equations are linear, any superposition of monochromatic waves is also a solution. Time-averaging a traveling wave over one cycle of the field (cycle-averaging) gives a smoothed intensity and energy density that evolve slowly and do not have sub-wavelength spatial resolution. For a traveling wave with sufficiently small frequency bandwidth, the electromagnetic energy flux given by the cycle-averaged Poynting vector $\overline{{\rm {\bf S}}({\rm {\bf x}},t)}$ is equal to the product of the cycle-averaged electromagnetic energy density $\overline{u({\rm {\bf x}},t)}$ and the energy-transport velocity $v_1 (\nu)$, so its cycle-averaged energy density is
\begin{equation}
\label{utoS}
\overline{u({\rm {\bf x}},t)}
= \left| \overline{{\rm {\bf S}}({\rm {\bf x}},t)} / {v_1 (\nu)} \right|
\approx \left| \overline{{\rm {\bf S}}({\rm {\bf x}},t)} /{v_g (\nu)} \right|,
\end{equation}
where the approximation requires a weakly dispersive medium to replace the energy-transport velocity $v_1 (\nu)$ with the group velocity,\cite{RN3987,RN477}
\begin{equation}
\label{eqvg}
v_g (\nu) = c /[ \partial (\nu n (\nu) )/ \partial \nu ].
\end{equation}
Cycle-averaging the Poynting vector in Eq. (\ref{tS}) and taking its scalar magnitude gives 
\begin{align}
\label{taStou}
\begin{split}
\overline{u(t)}
& = c
\left[ \dfrac{\epsilon (\nu)}{\mu (\nu)} \right]^{1/2}
\!
\dfrac{1}{2}({\rm {\bf E}}_{0} \cdot {\rm {\bf E}}_{0})
\left[ \dfrac {\partial (\nu n(\nu)) / \partial \nu} {c}\right] \\
& = \epsilon (\nu)
\dfrac{1}{2}({\rm {\bf E}}_{0} \cdot {\rm {\bf E}}_{0})
\left[ \dfrac {\partial (\nu n(\nu)) / \partial \nu} {n(\nu)}\right],
\end{split}
\end{align}
where the second line is obtained from the first line by inserting $(\epsilon \mu )^{1/2} / n = 1$. (The coordinate $\bf{x}$ has disappeared because cycle averaging a monochromatic traveling wave leaves no spatial resolution.) Equation (\ref{taStou}) matches Brillouin's Eq. (8), \cite{RN476} Pelzer's Eq. (5), \cite{RN3987} and Loudon's Eq. (19) \cite{RN3989,RN3990} for the cycle-averaged total electromagnetic energy per unit volume of a traveling quasi-monochromatic plane wave. Using $\epsilon \left|{\rm {\bf E}}_{0} \right|^2 = \mu \left|{\rm {\bf H}}_{0} \right|^2$ from Eqs. (\ref{HfromEvector})-(\ref{k}) and $n=(\epsilon \mu)^{1/2}$, Eq. (\ref{taStou}) can be recast into the same form as Eq. (80.12) of ref. \onlinecite{RN1985}, which reduces to the first line of Eq. (20) from Ch. IV of ref. \onlinecite{RN477} for non-magnetic media.

The spectral density for the square of the field $({{\rm {\bf E} \cdot {\bf E}}})^{\nu}$ can be accessed through the inverse Fourier transform of its time-domain correlation function in the same manner as the spectral density for the Poynting vector is accessed through Eqs. (\ref{taS}) - (\ref{sdS}), which gives
\begin{flalign}
\begin{split}
\label{sdE}
({{\rm {\bf E} \cdot {\bf E}}})^{\nu} ({\bf{x}},\nu) & \equiv \dfrac{\partial}{\partial \nu} 
\langle{{\rm {\bf E}}({\rm {\bf x}},t) \cdot
{\rm {\bf E}}({\rm {\bf x}},t)} \rangle _t  \\
& = \dfrac{1}{cycles} \langle{({\rm {\bf E}} \cdot
{\rm {\bf E}})({\bf{x}},\nu)} \rangle _t \:.
\end{split}
\end{flalign}
The functional dependence on position and frequency is written only once because the frequency-domain field for incoherent light has an indeterminate random-walk character\cite{RN3986,RN3985} while its time-averaged time-domain correlation function accumulates a stationary inverse Fourier transform.
The spectral density for electromagnetic energy per unit volume is
\begin{align}
\label{sdu}
\begin{split}
{u^{\nu}({\bf{x}},\nu)}
 &= \epsilon (\nu) ({{\rm {\bf E} \cdot {\bf E}}})^{\nu} ({\bf{x}},\nu)
\left[ \dfrac {\partial (\nu n(\nu)) / \partial \nu} {n(\nu)}\right].
\end{split}
\end{align}
\noindent Equation (\ref{sdI}) and Eq. (\ref{tS}) lead to the relationship
\begin{equation}
\label{sdEEsdI}
({{\rm {\bf E} \cdot {\bf E}}})^{\nu}({\rm{\bf x}}, \nu)
= \dfrac{n(\nu)}{c} \dfrac{1}{\epsilon(\nu)} I^{\nu}({\rm{\bf x}}, \nu).
\end{equation}
Combining Eqs. (\ref{sdI}), (\ref{utoS}), and (\ref{sdu}) relates the spectral irradiance to the product of the spectral density for electromagnetic energy per unit volume and the group velocity
\begin{equation}
\label{sdItosdu}
I^{\nu}({\bf{x}},\nu ) = u^{\nu}({\bf{x}},\nu ) v_g (\nu).
\end{equation}

Equations (\ref{sdu}) to (\ref{sdItosdu}) are equally valid between the positive-frequency spectral density for the square of the electric field, 
\begin{equation}
\label{pfsdE}
({{\rm {\bf E} \cdot {\bf E}}})^{\nu}_{+} ({\bf{x}},\nu) = 2 ({{\rm {\bf E} \cdot {\bf E}}})^{\nu} ({\bf{x}},\nu) \theta (\nu),
\end{equation}
the positive-frequency spectral density for electromagnetic energy per unit volume,
\begin{equation}
\label{pfsdu}
u^{\nu}_{+}({\bf{x}},\nu) =
2 u^{\nu}({\bf{x}},\nu) \theta (\nu),
\end{equation}
and the positive-frequency spectral density for the irradiance given by Eq. (\ref{pfsdI}). In an infinite, homogeneous, and isotropic medium, Eqs. (\ref{sdE}), (\ref{sdu}), (\ref{pfsdE}) and (\ref{pfsdu}) relate the spectral density for electromagnetic energy per unit volume to that for the square of the field
\begin{equation}
\label{sdu2sdE}
u^{\nu}_{+}({\bf{x}},\nu)
= \epsilon (\nu) ({{\rm {\bf E} \cdot {\bf E}}})^{\nu}_{+}({\bf{x}},\nu) \left[ \dfrac {\partial (\nu n(\nu)) / \partial \nu} {n(\nu)}\right].
\end{equation}

Since traveling waves have been cycle-averaged at Eq. (\ref{utoS}), Eqs. (\ref{sdu}) to (\ref{sdItosdu}) and  (\ref{sdu2sdE}) cannot provide sub-wavelength spatial resolution. Equation (\ref{sdu}) can be recast into the same form as Eq. (14) of ref. \onlinecite{RN4102} or into the different form mentioned below Eq. (23) of ref. \onlinecite{RN2514}; Huang's pioneering treatment of a model\cite{RN3988} for what are now known as phonon-polaritons\cite{RN4106} avoids the ambiguities from cycle averaging in ref. \onlinecite{RN477, RN3987, RN3989, RN3990, RN1985, RN2514, RN4102} to reveal the field-induced potential and kinetic energy of charge carriers that accompanies a monochromatic electromagnetic wave in a dispersive dielectric. Equation (\ref{sdu2sdE}) need not hold in a structured medium (for example, in a standing wave cavity, the electromagnetic energy of a mode can be wholly electric in one place and wholly magnetic in another).

Positive-frequency spectral densities for electromagnetic energy density per unit volume were used in the work of Planck\cite{RN2744} and of Einstein\cite{RN979} and have remained in use wherever only spectral densities for intensity or irradiance are needed (for example, absorption and fluorescence spectroscopy with wavelength resolving instruments\cite{RN1559}). Frequencies are usually restricted to be positive in quantum electrodynamics (see Eq. (2.13) in ref. \onlinecite{RN2982} ; Eq. (45) of complement A$_{I}$ in ref. \onlinecite{RN713} ; page 96 of ref. \onlinecite{RN3778}; or Eq. 13.11 in ref. \onlinecite{RN2315} ). On the other hand, both positive and negative frequencies are usually used when discussing the Kramers-Kronig dispersion relations between real and imaginary parts of the refractive index, permittivity, etc.,\cite{RN1580,RN1912,RN3778,RN2558} in Fourier transform spectroscopy,\cite{RN311,RN1013} and in nonlinear optics.\cite{ RN398,RN524,RN2415,RN3129} Appendix D shows that the factor of 2 in Eq. (\ref{pfsdu}) is necessary to connect the Einstein coefficients (which use the positive-frequency spectral density for electromagnetic energy per unit volume) to transition probabilities from quantum mechanical first-order time-dependent perturbation theory with a classical field (where a complex-valued Fourier transform arises naturally from the complex exponential phase factors in the time-dependent Schr\"{o}dinger equation and generates a frequency-domain field with both positive and negative frequencies).

Taking the expectation value of the electromagnetic energy density by inserting the electric field operator for a single quantized field mode $(\mbox{\boldmath$\varepsilon$},\textbf{k})$ from Eq. (\ref{eq22}) into Eq. (\ref{sdu}) and Eq. (\ref{pfsdu}) and using the matrix elements in Eq. (\ref{eq76}) yields the quantum electromagnetic energy density; 
\begin{flalign}
\begin{split}
\label{qed}
\langle \tilde{u}({\mbox{\boldmath$\varepsilon$},\textbf{k} }) \rangle & = \sum\nolimits_{l=0}^\infty  
{ \mathcal{P}_{{\mbox{\footnotesize \boldmath$\varepsilon$}},{\rm {\bf k}}} }(l_{{\mbox{\footnotesize \boldmath$\varepsilon$}},{\rm {\bf k}}}) {h\nu_{\bf k}} (l_{{\mbox{\footnotesize \boldmath$\varepsilon$}},{\rm {\bf k}}} + 1/2)/V \\
& = u({\mbox{\boldmath$\varepsilon$},\textbf{k} }) + (1/2){h\nu_{\bf k}}/V,
\end{split}
\end{flalign}
where $\mathcal{P}_{\mbox{\footnotesize \boldmath$\varepsilon$},{\rm {\bf k}}} (l_{{\mbox{\footnotesize \boldmath$\varepsilon$}},{\rm {\bf k}}})$ is the probability that $l_{\mbox{\boldmath$\varepsilon$},\textbf{k} }$ is the photon number in mode $(\mbox{\boldmath$\varepsilon$},\textbf{k})$,  $u({\mbox{\boldmath$\varepsilon$},\textbf{k} })$ is the mode's classical electromagnetic energy density from Eq. (\ref{eq9}), and $(1/2)h\nu_{\bf{k}}$ is the mode's quantum zero-point energy.

For a single coherent pulse of light, its complex-valued frequency-domain spectrum of the classical field is obtained from the time-domain field by inverse Fourier transformation:
\begin{equation}
\label{eqA6}
\mathbf{\hat{{{\cal E}}}}({\rm {\bf x}},\nu )=\int_{-\infty }^{+\infty } 
{{\rm {\bf E}}({\rm {\bf x}},t)\exp (i2\pi _c \nu t)dt} .
\end{equation}

\noindent The dimensions of the frequency-domain electric field are $\dim [\mathbf{\hat{{{\cal E}}}}({\rm {\bf x}},\nu )]=(mass^{1/2}/length^{1/2})$. The frequency in Eq. (\ref{eqA6}) can be either positive or negative. Because the time-domain fields are real-valued, Eq. (\ref{eqA6}) gives a complex-conjugate symmetry to the frequency-domain fields:
\begin{equation}
\label{ar3}
\mathbf{\hat{{{\cal E}}}}^{\ast }({\rm {\bf x}},\nu )=\mathbf{\hat{{{\cal 
E}}}}({\rm {\bf x}},-\nu ).
\end{equation}

\noindent By Rayleigh's theorem,\cite{RN447} 
\begin{flalign}
\begin{split}
\label{ar4}
& \int_{-\infty }^{+\infty } {{\rm {\bf E}}({\rm {\bf x}},t)\cdot {\rm {\bf 
E}}({\rm {\bf x}},t)dt} \\ 
& \quad \quad =\dfrac{1}{cycles}\int_{-\infty }^{+\infty } 
{\mathbf{\hat{{{\cal E}}}}^{\ast }({\rm {\bf x}},\nu )\cdot 
\mathbf{\hat{{{\cal E}}}}({\rm {\bf x}},\nu )d\nu } ,
\end{split}
\end{flalign}
\noindent Rayleigh's theorem is the zero-lag case of the correlation theorem, so the frequency integral is a zero-lag Fourier transform and has the same $1/cycles$ prefactor from $\dim[\nu]=cycles/time$. Unlike the above power spectrum of a pulse, the time-averaged spectral density of electromagnetic energy per unit volume [Eq. (\ref{sdu2sdE})] that appears in the Einstein relations vanishes for a single pulse as the duration of the time average in Eq. (\ref{tcS}) goes to infinity. This is connected to the difference between the quantum mechanical calculation for one event \textit{vs.} the equilibrium time-averaged rate for many events.

Gouy's equivalent representations of partially coherent light as monochromatic waves with random phases or full-spectrum pulses with random timing\cite{RN1274} can be used to connect the power spectrum of a pulse to the spectral density of electromagnetic energy per unit volume for an infinite sequence of randomly timed full-spectrum pulses. Gouy's two representations are equivalent as far as the first-order correlation functions and first-order coherence\cite{RN424,RN3,RN4003} that appear in linear spectroscopy. In fact, we can replace a sequence of randomly timed full-spectrum pulses with a randomly timed sequence of different partial-spectrum pulses so long as the average spectral density is the same over the narrowest frequency intervals of interest. (Smooth spectra require an ensemble average on top of the long-time average.\cite{RN3986}) 

Viewing incoherent light as a sequence of randomly timed identical pulses,\cite{RN1274} the spectral density of the square of the field in Eq. (\ref{sdu}) can be obtained by multiplying the power spectrum of a single pulse from Rayleigh's theorem [Eq. (\ref{ar4})] by the average pulse repetition rate $k_{pulses}$ ($\dim [k_{pulses}] = 1/time$) for the randomly timed sequence and converting to the spectral density using Eq. (\ref{sdE}),
\begin{equation}
\label{sdscp}
({{\rm {\bf E} \cdot {\bf E}}})^{\nu} = \dfrac {1}{cycles}
\left[ \mathbf{\hat{{{\cal E}}}}^{\ast }({\rm {\bf x}},\nu )\cdot 
\mathbf{\hat{{{\cal E}}}}({\rm {\bf x}},\nu ) \right] k_{pulses}.
\end{equation}
The positive-frequency spectral density of electromagnetic energy per unit volume is obtained by substituting Eq. (\ref{sdscp}) into Eqs. (\ref{sdu}) and (\ref{pfsdu}). Accounting for the contributions from magnetic and mechanical energy densities\cite{RN3988,RN3987} gives
\begin{align}
\label{pfsducp}
\begin{split}
{u_{+}^{\nu}({\rm {\bf x}},\nu)}
&= \dfrac{1}{cycles} 2 \epsilon (\nu) \left[ \dfrac {\partial (\nu n(\nu))/ \partial \nu} {n(\nu)}\right] \\
& \quad \quad \cdot \mathbf{\hat{{{\cal E}}}}^{\ast }({\rm {\bf x}},\nu ) \cdot \mathbf{\hat{{{\cal E}}}}({\rm {\bf x}},\nu ) k_{pulses} \theta (\nu).
\end{split}
\end{align}
\noindent For weakly dispersive transparent media, the positive-frequency spectral density for the irradiance can be obtained by multiplying the spectral density for the electromagnetic energy per unit volume with the group velocity $v_{g} (\nu)$ in Eq. (\ref{eqvg}),
\begin{align}
\label{pfsdIcp}
\begin{split}
{I_{+}^{\nu}({\rm {\bf x}},\nu)}
&= \dfrac{1}{cycles} 2  \left[ \dfrac {\epsilon (\nu) c} {n(\nu)}\right] \\
& \quad \quad \cdot \mathbf{\hat{{{\cal E}}}}^{\ast }({\rm {\bf x}},\nu ) \cdot \mathbf{\hat{{{\cal E}}}}({\rm {\bf x}},\nu ) k_{pulses} \theta (\nu).
\end{split}
\end{align}

\noindent The relationships in Eqs. (\ref{pfsducp}) and (\ref{pfsdIcp}) can be used to connect the transition probabilities from quantum mechanical time-dependent perturbation theory with a classical field to the Einstein $B$-coefficient spectra.

\renewcommand{\theequation}{B-\arabic{equation}}
\setcounter{equation}{0}
\section*{Appendix B}
This appendix proves that if a pair of functions obeys the generalized Einstein relation between stimulated transition spectra in Eq. (\ref{eq14b}), then convolution with corresponding functions from any pair of functions that obey the stimulated transition generalized Einstein relation generates a new pair of functions that also obey the generalized Einstein relation for stimulated transitions. In the proof, we subsume factors of $\exp (\Delta \mu ^{\rm{o}} / k_{\rm{B}}T)$ into the forward stimulated functions and let $x = k_{\rm{B}}T / h$.  Specifically,
let
\begin{subequations}
\label{eqB1}
\begin{equation}
\label{eqB1a}
g(-\nu)=f(\nu) \exp(-\nu /x).
\end{equation}
be the pair that obeys Eq. (\ref{eq14b}), and
\begin{equation}
\label{eqB1b}
q(-\nu)=p(\nu) \exp (-\nu /x),
\end{equation}
\end{subequations}
be the corresponding pair they are convolved with. If
\begin{subequations}
\label{eqB2}
\begin{equation}
\label{eqB2a}
G(\nu)= \int_{-\infty}^{+\infty} {g(y)q(\nu -y) dy},
\end{equation}
and
\begin{equation}
\label{eqB2b}
F(\nu)= \int_{-\infty}^{+\infty} {f(y)p(\nu -y) dy},
\end{equation}
\end{subequations}
give the convolutions, then
\begin{equation}
\label{eqB3}
G(-\nu)=F(\nu) \exp(-\nu /x).
\end{equation}

\noindent PROOF: Reverse the sign of $\nu$ everywhere in Eq. (\ref{eqB2a}) for $G(\nu)$, substitute for $g(y)$ using Eq. (\ref{eqB1a}) with sign-reversed arguments, substitute for $q(-\nu - y)$ using Eq. (\ref{eqB1b}), combine the exponentials and pull $\exp(-\nu/x)$ outside the integral, substitute $z=-y$ inside the integral, and use Eq. (\ref{eqB2b}) for $F(\nu)$ to obtain Eq. (\ref{eqB3}).
\begin{flalign}
\begin{split}
\label{eqB4}
G(-\nu) &= \int_{-\infty}^{+\infty} {g(y) q(-\nu -y) dy}, \\
&= \int_{-\infty}^{+\infty} {f(-y) \exp (y/x) q(-\nu -y) dy}, \\
&= \int_{-\infty}^{+\infty} {f(-y)\exp (y/x) p(\nu + y) \exp[-(\nu + y)/x]dy} \\
&= \exp(-\nu/x) \int_{-\infty}^{+\infty} {f(-y) p(\nu + y)  dy} \\
&= \exp(-\nu/x) \int_{-\infty}^{+\infty} {f(z) p(\nu - z)  dz} \\
&= F(\nu)\exp(-\nu/x).
\end{split}
\end{flalign}

\noindent COROLLARY:  If functions $f$ and $g$ satisfy Eq. (\ref{eqB1}), functions $F$ and $G$ are obtained from them by linear functionals satisfying
\begin{subequations}
\label{eqB5}
\begin{equation}
\label{eqB5a}
G(\nu)= \int_{-\infty}^{+\infty} {g(y)Q(\nu ; -y) dy},
\end{equation}
\begin{equation}
\label{eqB5b}
\quad \quad \quad F(\nu)= \int_{-\infty}^{+\infty} {f(y)P(\nu ; -y) dy},
\end{equation}
\end{subequations}
and
\begin{equation}
\label{eqB5c}
Q(-\nu ; -y) = P(\nu ; -y) \exp [-(\nu +y)/x],
\end{equation}
then Eq. (\ref{eqB3}) holds.

The proof follows the same steps as in Eq. (\ref{eqB4}), with the functionals $P$ and $Q$ in place of the functions $p$ and $q$, respectively. The functional relationships in Eq. (\ref{eqB5}) and (\ref{eqB5c}) allow $Q(-\nu ; -y)$ to be replaced by $Q'(-\nu; +y)$ in both; this change of sign for $y$ alone would not be possible with convolution.

In consequence of this corollary, each frequency component of a pair of spectra $f$ and $g$ that obey the generalized Einstein relation can be convolved with a different pair of corresponding lineshapes $P$ and $Q'$ that obey the generalized Einstein relation and depend on the frequency as a parameter, and the result of adding these ``pseudo-convolved'' spectra will still obey the generalized Einstein relation. Equality (\ref{rGER}) is a generalization of Eq. (\ref{eqB5c}) in the $Q'(-\nu; +y)$ form with
\begin{subequations}
\label{eqB7}
\begin{equation}
\label{eqB7a}
y = (E_{S \eta '} -E_{R \eta})/h,
\end{equation}
\begin{equation}
\label{eqB7b}
Q'(-\nu; +y)= {r_{SE'\eta',RE\eta}(E_{S \eta '} -E_{R \eta } - h\nu )},
\end{equation}
\begin{equation}
\label{eqB7c}
P(\nu; -y) ={r_{RE\eta,SE'\eta'}(E_{R \eta} -E_{S \eta '} + h\nu )}.
\end{equation}
\end{subequations}
The generalization is that the pair of functionals $r$ can depend on both energies $E$ and $E'$ plus both configurations $\eta$ and $\eta'$ and not just on the energy difference in the Bohr frequency $y$. With this generalization indicated by the subscripts on the pair $r$, the single arguments in $r$ are not a restriction over the two arguments in the functionals $Q'$ and $P$. If the reverse and forward functionals $r$ obey Eq. (\ref{rGER}) (as opposed to some generalization which might include changes in chemical potential from a configuration-dependent interaction of the molecule + environment with the radiation field and thermal reservoir) then the reverse and forward $B$-coefficient spectra will obey the generalized Einstein relation in Eq. (\ref{eq21}) with the change in standard chemical potential calculated from the quantum levels of molecule + environment as in Eq. (\ref{eqP}). In this case, with a sufficiently large environment, the configuration-dependent effects of the molecule + environment interaction with the radiation field and thermal reservoir that are incorporated into $r$ should arise from the radiation field because its interaction with the molecule is not reduced by larger and larger environments between the molecule and the thermal reservoir. We will assume this to be the case because the environment and thermal reservoir must be large relative to their interface in order to define temperature (see Gibbs' discussion about neglect of their interaction on page 37 of ref. \onlinecite{RN3991}).

\renewcommand{\theequation}{C-\arabic{equation}}
\setcounter{equation}{0}
\section*{Appendix C}
In this appendix, we illustrate how the quantities in Einstein-coefficient spectra reduce to those in Einstein coefficients for line spectra in vacuum, bringing out a difference between the definition of the square of the transition dipole moment between levels used here and that used in ref. \onlinecite{RN1457}. The initial upper and final lower \textit{levels} are indicated by lowercase italic $s$ and $r$, respectively. For the electric-dipole matrix element in Eq. (\ref{LFeq}), the band labels $S$ and $R$ are replaced by the level labels $s$ and $r$, the continuous energies $E '$ and $E$ are no longer needed because they are already fully specified by the levels, and the continuous configuration variables $\eta '$ and $\eta $ are replaced by complete sets of discrete angular momentum projection quantum numbers $m'$ and $m$ needed to fully specify eigenstates within degenerate levels $s$ and $r$, respectively. MKSA units are used for comparison to ref. \onlinecite{RN1457} ($\epsilon_0 = 1$ in rationalized Heaviside-Lorentz units\cite{RN1580,RN2982}).

For line spectra, in the absence of the radiation-matter interaction, we have the initial conditional probability distribution
\begin{equation}
P_{s}^{E}(E',m') = \delta ^{E} \! (E_s - E') p(m'|s),
\end{equation}
where $p(m'|s)$ is the conditional probability of $m'$ given $s$ [so we have $\sum_{m'} p(m'|s) = 1$], and the density of final states is
\begin{equation}
    \rho _{r}^{E}(E,m) = \delta ^{E} \! (E_r - E).
\end{equation}
(There is one final eigenstate for each complete set of final quantum numbers $m$.) The temperature-dependent radiation-matter interaction broadens the spectrum
\begin{equation}
r_{sE'm',rEm}(E' - E - h\nu ),
\end{equation}
so that the normalized lineshape is
\begin{flalign}
\begin{split}
\label{evalINT}
& r_{sE'm',rEm}(E_{s} -E_{r} - h\nu ) \\
& \quad \quad =\sum_{m'} \int_{E' , E} 
 P_{s}^{E}(E',m')
\cdot \rho_{r}^{E} (E,m) \\
& \quad \quad \quad \quad \quad \cdot r_{sE'm',rEm}(E' -E - h\nu ) dE dE' ,
\end{split}
\end{flalign}
and the dipole-strength spectrum from Eq. (\ref{eq27}) can be written
\begin{flalign}
\begin{split}
\label{d2vac}
&d_{s\to r}^{2} (\nu, T) = \sum\limits_{m}
\left\langle{ \left| {\rm {\bf {d}}}_{sm',rm} \right| ^2 }\right\rangle _{m'} \\
& \quad \quad \quad \quad \quad \quad \cdot
r_{sE'm',rEm}(E_{s} - E_{r} - h\nu ) . 
\end{split}
\end{flalign}
\noindent At equilibrium, $p(m'|s)=1/g_s$, where $g_s$ is the degeneracy of level $s$. Degenerate states within an isolated energy level only arise if required by symmetry.\cite{RN1984} With an isotropic average over excitation polarization, symmetry requires that all degenerate initial states have equal total outgoing conditional transition probabilities.\cite{RN725}

The isotropic average over $m'$ in Eq. (\ref{d2vac}) is equivalent to
\begin{flalign}
\begin{split}
\label{vecd2}
 \left\langle \left| {\rm {\bf {d}}}_{sm',rm} \right| ^2  \right\rangle_{m'} & =\dfrac{1}{g_{s} 
}\sum\limits_{m'} {\left| {\rm {\bf {d}}}_{sm',rm} \right| ^2 }\\ 
& =\left\langle {\dfrac{1}{g_{s} }\sum\limits_{m'} {\left| {\rm {\bf {d}}}_{sm',rm} \right| ^2 } }
\right\rangle_{m} \\ 
& =\left\langle {\left| {\rm {\bf {d}}}_{sm',rm} \right| ^2 } \right\rangle_{m} \equiv 
\left| {{\rm {\bf d}}_{sr} } \right|^{2}, \\ 
\end{split}
\end{flalign}
\noindent where the second equality holds because the sum over $m'$ is independent of $m$ and the third equality holds because the average over $m$ is independent of $m'$. The equality between averages over $m$ and $m'$ justifies the definition of $\left| {{\rm {\bf d}}_{sr} } \right|^{2}$ as the square of the transition-dipole moment - Eq. (\ref{vecd2}) corresponds to Eq. (\ref{eq26}) in this paper [which defines an average over three perpendicular polarization directions that multiplies their sum by a factor of (1/3)] but differs significantly from the definition in Eq. (29) of ref. \onlinecite{ RN1457}. Unlike Eq. (29) of ref. \onlinecite{ RN1457}, the definition of the square of the transition-dipole moment chosen here is independent of the direction ($r \to s$ \textit{vs.} $s \to r$) of the transition.

The total emission dipole strength in Eq. (\ref{eq31}) is
\begin{flalign}
\begin{split}
\label{D2vac}
D_{s\to r}^{2} =\sum\limits_m {\left\langle {\left| {\left\langle {s,m'} 
\right|{\rm {\bf \hat{d}}}\left| {r,m} \right\rangle } \right|^{2}} 
\right\rangle_{m'} } =\left| {{\rm {\bf d}}_{sr} } \right|^{2} \cdot g_{r} ,
\end{split}
\end{flalign}
\noindent where the sum over final state quantum numbers $m$ gives the degeneracy of the final level. The total emission dipole strength defined here is equal to the ``square of the transition dipole moment'' as defined in Eq. (29) of ref. \onlinecite{RN1457} and depends on the direction of the transition. In vacuum, Eq. (\ref{eq28}) and the  integration over frequency in Eq. (\ref{INTABb}) yield
\begin{equation}
\label{BDem}
B_{s\to r} =\dfrac{D_{s\to r}^{2} }{6\epsilon_{0} \hslash^{2}}.
\end{equation}
Eqs. (\ref{vecd2}) and (\ref{D2vac}) show that this implies
\begin{equation}
\label{BDmix}
B_{r\to s} =\dfrac{g_{s} }{g_{r} }\dfrac{D_{s\to r}^{2} }{6\epsilon 
_{0} \hslash^{2}}.
\end{equation}
Equation (\ref{BDmix}) is precisely the relationship in Table I of ref. 
\onlinecite{RN1457}. Equation (\ref{eq34}) shows 
it could also be written as
\begin{equation}
\label{BDabs}
B_{r\to s} =\frac{D_{r\to s}^{2} }{6\epsilon_{0} \hslash^{2}}.
\end{equation}

The alternative forms
\begin{equation}
\label{BDvecd2}
B_{r\to s} =\dfrac{ \left| {{\rm {\bf d}}_{rs}  } 
\right|^{2} g_{s}}{6\epsilon_{0} \hslash^{2}}
\quad \text{and} \quad
B_{s\to r} 
=\dfrac{\left| {{\rm {\bf d}}_{sr} } \right|^{2} g_{r} }{6\epsilon_{0} 
\hslash^{2}}
\end{equation}
with $|{{\rm {\bf d}}_{rs}}|^{2} = |{{\rm {\bf d}}_{sr}}|^{2}$ more clearly illuminate the role of the final state degeneracy in the Einstein $B$-coefficients and [through Eqs. (\ref{BDem}) and (\ref{BDabs})] the total dipole strengths.

\renewcommand{\theequation}{D-\arabic{equation}}
\setcounter{equation}{0}
\section*{Appendix D}

This appendix derives the transition probability per unit time from the transition probabilities for a single initial quantum state under the action of a single short pulse by using quantum mechanical time-dependent perturbation theory with a classical field. The derivation of the transition probability highlights the effect of using the positive-frequency spectral density for electromagnetic energy per unit volume in defining the Einstein coefficients. The relationship between the transition probability from quantum mechanical first-order time-dependent perturbation theory and the conditional transition probability per unit time depends on the relationship between the complex-valued frequency-domain classical electric field, which is spread over both positive and negative frequencies, and the positive-frequency spectral density for electromagnetic energy per unit volume in Eq. (\ref{pfsducp}). Since eigenstates with degeneracies are used, the formulas are only given for vacuum. The notation for states and levels used here is specified in the first paragraph of Appendix C. 

For time-dependent perturbation theory, the time-dependent molecular wavefunction is expanded in the complete basis of eigenstates of the field-free Hamiltonian:
\begin{equation}
\left| \psi (t) \right\rangle =\sum\limits_n {c_{n} (t)\exp (-iE_{n} t/\hslash 
)\left| n \right\rangle } 
\end{equation}
where the field-free eigenstates satisfy
\[
\hat{{H}}_{0} \left| n \right\rangle =E_{n} \left| n \right\rangle ,
\]
and the complex-valued basis-state coefficients are normalized
\[
\sum\limits_n {\left| {c_{n} (t)} \right|}^{2}=1,
\]
and remain constant in the absence of an external field. Adding the electric 
dipole interaction with a classical time-dependent electric field to the 
field-free Hamiltonian,
\begin{equation}
\label{semi}
\hat{{H}}=\hat{{H}}_{0} -{\rm {\bf \hat{{d}}}}\cdot {\rm {\bf E}}(t)
\end{equation}

\noindent where ${\rm {\bf \hat{d}}}$ is the electric-dipole operator, the position dependence of the field has been suppressed, and the wavefunction's basis-state coefficients become functions of time while the field is on. The time-dependent wavefunction coefficients are
\begin{flalign}
\begin{split}
c_{rm} (t) & = c_{rm} (0)\\
& +\dfrac{i}{\hslash }\sum\limits_{sm'} \int_0^t {c_{sm'} (t')} \\
& \quad \quad \quad \quad \quad {\rm {\bf d}}_{rm,sm'} \cdot {\rm {\bf E}}(t')\exp (i2\pi _c \nu_{s\to r} t')dt' ,
\end{split}
\end{flalign}

\noindent where 
${\rm {\bf d}}_{rm,sm'} =\left\langle {r,m} \right|{\rm {\bf \hat{{d}}}}
\left| {s,m'}\right\rangle $ are laboratory-frame vector electric-dipole matrix elements and $\nu_{s\to r} =(E_{r} -E_{s} )/h$ is the (negative) Bohr frequency arising from the time-dependent Schr\"{o}dinger equation for the stimulated emission transition from $s$ to $r$. (The use of laboratory-frame vector electric-dipole matrix elements between fully specified eigenstates here allows the dot product with the electric field to be pulled outside the rotational eigenfunctions.) The first-order time-dependent perturbation theory 
approximation is
\begin{flalign}
\begin{split}
c_{rm} (t) & \approx c_{rm} (0) \\
& +\dfrac{i}{\hslash }\sum\limits_{sm'} \int_0^t {c_{sm'} (0)} \\
& \quad \quad \quad \quad \quad {\rm {\bf d}}_{rm,sm'} \cdot {\rm {\bf E}}(t')\exp 
(i2\pi _c \nu_{s\to r} t')dt' .
\end{split}
\end{flalign}

If the pulse of light is weak [so that the integral is small compared to $c_{sm'} (0)$] and confined to the time interval between 0 and $t$, then both limits of the time integral may be extended to infinity to get a first-order result that is valid after the pulse of light is over.

\begin{flalign}
\begin{split}
c_{rm} (t_{+} ) & \approx c_{rm} (0) \\
& +\dfrac{i}{\hslash }\sum\limits_{sm'} 
\int_{-\infty }^{+\infty } {c_{sm'} (0)}\\
& \quad \quad \quad \quad \quad {\rm {\bf d}}_{rm,sm'} 
\cdot {\rm {\bf E}}(t)\exp (i2\pi _c \nu_{s\to r} t)dt ,
\end{split}
\end{flalign}

\noindent where $t_{+} $ indicates any time after the pulse is over. This extension of the integration limits has converted the time integral into an inverse 
Fourier transform of the time-domain field given by Eq. (\ref{eqA6})
\begin{equation}
c_{rm} (t_{+} )\approx c_{rm} (0)+\dfrac{i}{\hslash }\sum\limits_{sm'} {c_{sm'} (0)
{\rm {\bf d}}_{rm,sm'} \cdot \mathbf{\hat{{{\cal 
E}}}}(\nu_{s\to r} )}
\end{equation}

The exact Bohr-frequency in the frequency-domain field is only compatible with the time-energy uncertainty principle because of this infinite extension of the time integration limits. By Eq. (\ref{ar3}), the complex-valued classical frequency domain electric field obeys $\mathbf{\hat{{{\cal E}}}}(-\nu )=\mathbf{\hat{{{\cal E}}}}^{\ast }(\nu )$. For comparison to the Einstein coefficients, we are interested in the transition probability starting from a wavefunction that is composed entirely of a single eigenstate. Choosing $\vert {\kern 
1pt}\,c_{sm'} (0){\kern 1pt}\vert \;=1$, the first-order time-dependent perturbation theory result for the final lower state coefficients is
\begin{equation}
c_{rm} (t_{+} )=\dfrac{i}{\hslash }c_{sm'} (0) {\rm {\bf 
d}}_{rm,sm'} \cdot \mathbf{\hat{{{\cal E}}}}(\nu_{s\to r} )
\end{equation}

The transition probabilities are a second-order perturbation theory result, but correctly given by the coefficients from first-order perturbation theory in the absence of coherence between initial states. The total conditional transition probability from one initial eigenstate \textit{sm'} to one final eigenstate \textit{rm} arising from a single pulse is
\begin{equation}
\mathcal{P}_{sm'\to rm} =\dfrac{\left| {c_{rm} (t_{+} )} \right|^{2}}{\left| {c_{sm'} 
(0)} \right|^{2}}=\dfrac{
\left| {{\rm {\bf d}}_{rm,sm'} \cdot 
\mathbf{\hat{{{\cal E}}}}(\nu_{s\to r} )} \right|^{2}}{\hslash^{2}}\quad .
\end{equation}

\noindent Using the same methods as Appendix C, this may be isotropically averaged over polarization, averaged over degenerate initial eigenstates $sm'$ within the initial level $s$, summed over degenerate final eigenstates \textit{rm} within the final level $r$, and rewritten as
\begin{equation}
\label{spr}
\sum\limits_m {\left\langle {\mathcal{P}_{sm'\to rm} } \right\rangle_{m'} } =g_{r} 
\dfrac{
\left| {{\rm {\bf d}}_{rs} } \right|^{2}\left| 
{\mathbf{\hat{{{\cal E}}}}(\nu_{s\to r} )} \right|^{2}}{3\hslash^{2}}\quad ,
\end{equation}

\noindent where $g_{r} $ is the degeneracy of the final level $r$. The squared magnitude of the field in Eq. (\ref{spr}) is an even function that extends over the entire frequency axis, with its power symmetrically distributed over both positive and negative frequencies. 

The total transition probability for a single pulse must be multiplied by the average rate at which pulses arrive to give the conditional transition probability per unit time,
\begin{equation}
\label{avG}
^{b}\Gamma_{s\to r} =\sum\limits_m {\left\langle {\mathcal{P}_{sm'\to rm} } \right\rangle 
_{m'} } k_{pulses} ,
\end{equation}
appearing in Eq. (\ref{eq2a}). Inserting Eq. (\ref{spr}) into Eq. (\ref{avG}) gives
\begin{flalign}
\begin{split}
\label{avGEk}
& ^{b}\Gamma_{s\to r} =
\dfrac{
\left| {{\rm {\bf d}}_{sr} } \right|^{2} g_{r} \left| 
{\mathbf{\hat{{{\cal E}}}}(\nu_{s\to r} )} \right|^{2}}{3\hslash^{2}} k_{pulses},
\end{split}
\end{flalign}
The next step is to rewrite the squared magnitude of the field using the positive-frequency spectral density of electromagnetic energy per unit volume in Eq. (\ref{pfsducp}), which gives (in MKSA units)
\begin{flalign}
\begin{split}
\label{avGu}
& ^{b}\Gamma_{s\to r} = (cycles)
\dfrac{\left| {{\rm {\bf d}}_{sr} } \right|^{2} g_{r}}
{6 \epsilon_0 \hslash^{2}} 
u_{+}^{\nu}(|\nu_{s\to r}|),
\end{split}
\end{flalign}
the discrete spectrum vacuum analog of Eq. (\ref{eq28}). Note that the 3 in the denominator of Eq. (\ref{avGEk}) became a 6 in the denominator of Eq. (\ref{avGu}) because the positive-frequency spectral density for electromagnetic energy per unit volume is a factor of 2 larger than the even frequency-domain spectral density [Eq. (\ref{pfsducp})].  Comparison of Eq. (\ref{eq2a}) with Eq. (\ref{avGu}) gives the vacuum Einstein $B$-coefficient in Eq. (\ref{BDvecd2}). The discrete-continuous spectrum relationships in Appendix C show how to extend the above derivation to dispersive and approximately transparent dielectric media, in which it yields Eq. (\ref{eq28}).

\section*{NOTES AND REFERENCES}
\bibliography{EDgerJCP}
\end{document}